\documentclass[fleqn,usenatbib]{mnras}

\usepackage[T1]{fontenc}

\DeclareRobustCommand{\VAN}[3]{#2}
\let\VANthebibliography\thebibliography
\def\thebibliography{\DeclareRobustCommand{\VAN}[3]{##3}\VANthebibliography}

\usepackage{graphicx}	
\usepackage{amsmath}	
\usepackage{amssymb}	
\usepackage{tabu}
\usepackage{longtable}
\usepackage{makecell}
\usepackage{multicol,multirow}
\usepackage{xcolor}
\usepackage{ulem}

\usepackage{newtxtext,newtxmath}

\title[Galaxy interactions trigger local type 2 quasars]{Galaxy interactions are the dominant trigger for local type 2 quasars}

\author[J. C. S. Pierce et al.]{J. C. S. Pierce,$^{1}$\thanks{E-mail: j.pierce3@herts.ac.uk}
C. Tadhunter,$^{2}$
C. Ramos Almeida,$^{3,4}$
P. Bessiere,$^{3,4}$
J. V. Heaton,$^{2}$
S. L. Ellison,$^{5}$
\newauthor
G. Speranza,$^{3,4}$
Y. Gordon,$^{6}$
C. O'Dea,$^{7}$
L. Grimmett$^{2}$
and L. Makrygianni$^{8}$
\\ \\
$^{1}$Centre for Astrophysics Research, University of Hertfordshire, College Lane, Hatfield AL10 9AB, UK\\
$^{2}$Department of Physics \& Astronomy, University of Sheffield, Sheffield S3 7RH, UK\\
$^{3}$Instituto de Astrof\'isica de Canarias, Calle V\'ia L\'actea,  s/n, E-38205 La Laguna, Tenerife, Spain\\
$^{4}$Departamento de Astrof\'isica, Universidad de La Laguna, E-38205 La Laguna, Tenerife, Spain \\
$^{5}$Department of Physics \& Astronomy, University of Victoria, Finnerty Road, Victoria, British Columbia V8P 1A1, Canada\\
$^{6}$Department of Physics, University of Wisconsin-Madison, 1150 University Ave, Madison, WI 53706, USA\\
$^{7}$Department of Physics \& Astronomy, University of Manitoba, Winnipeg, MB R3T 2N2, Canada\\
$^{8}$The School of Physics \& Astronomy, Tel Aviv University, Tel Aviv 69978, Israel
}

\date{Accepted XXX. Received YYY; in original form ZZZ}

\pubyear{2022}

\begin{document}
\label{firstpage}
\pagerange{\pageref{firstpage}--\pageref{lastpage}}
\maketitle

\begin{abstract}
The triggering mechanism for the most luminous, quasar-like active galactic nuclei (AGN) remains a source of debate, with some studies favouring triggering via galaxy mergers, but others finding little evidence to support this mechanism. 
Here, we present deep Isaac Newton Telescope/Wide Field Camera imaging observations of a complete sample of 48 optically-selected type 2 quasars -- the QSOFEED sample (L$_{\rm [OIII]}>$10$^{8.5}L_{\sun}$; $z < 0.14$).
Based on visual inspection by eight classifiers, we find clear evidence that galaxy interactions are the dominant triggering mechanism for quasar activity in the local universe, with 65$^{+6}_{-7}$ per cent of the type 2 quasar hosts showing morphological features consistent with galaxy mergers or encounters, compared with only 22$^{+5}_{-4}$ per cent of a stellar-mass- and redshift-matched comparison sample of non-AGN galaxies -- a 5$\sigma$ difference. The type 2 quasar hosts are a factor 3.0$^{+0.5}_{-0.8}$ more likely to be morphologically disturbed than their matched non-AGN counterparts, similar to our previous results for powerful 3CR radio AGN of comparable [OIII] emission-line luminosity and redshift.
In contrast to the idea that quasars are triggered at the peaks of galaxy mergers as the two nuclei coalesce, and only become visible post-coalescence, the majority of morphologically-disturbed type 2 quasar sources in our sample are observed in the pre-coalescence phase (61$^{+8}_{-9}$ per cent). We argue that much of the apparent ambiguity that surrounds observational results in this field is a result of differences in the surface brightness depths of the observations, combined with the effects of cosmological surface brightness dimming.
\end{abstract}




\section{Introduction}
\label{sec:introduction}

Quasars have long drawn attention as the most luminous AGN ($L_{\rm bol} > 10^{38}$\,W; $L_{\rm[OIII]} > 10^{35}$\,W). Over the six decades since their discovery, there has been increasing recognition that they are not merely by-products of the growth of galaxies and their super-massive black holes (SMBHs) by gas accretion, but that they may directly affect this growth. Observations show that their powerful winds and jets heat, ionize and eject the gas from the circum-nuclear regions, thereby affecting  the star-formation histories of the host galaxies \citep[e.g.][]{fab12,veilleux13,harrison17}. At the same time, the incorporation of such feedback effects into models of galaxy evolution allows certain properties of the general galaxy population to be explained, including the correlations between SMBH mass and host galaxy properties \citep[e.g.][]{silk98,dm05,kormendy13}, and the shape of the high luminosity end of the galaxy luminosity function \citep[e.g.][]{cro06,bow06}.  Therefore, it is important to understand how quasars relate to the general galaxy population and are triggered as galaxies evolve.

To appear as a quasar, even at the lower end of the quasar luminosity range, the SMBH in a galaxy must accrete gas at a rate  $\dot{M}> 0.2$\,M$_{\odot}$\,yr$^{-1}$ for a typical efficiency of $\eta\sim0.1$, and this accretion rate must be sustained over a typical quasar lifetime of $\sim$$10^6$ -- $10^8$\,yr \citep[e.g.][]{martini01,martini04,hopkins05}.  The quasar triggering problem, therefore, amounts to understanding how the required total of at least $\sim$$10^5$ -- $10^7$\,M$_{\odot}$ of gas can lose sufficient angular momentum to be transported from kiloparsec scales to sub-parsec scales, so that it can be accreted by the SMBH.

From an observational perspective, investigating quasar triggering mechanisms is challenging for two main reasons. First, in order to directly detect the gas flows that trigger the activity it is important to resolve spatial scales around the galactic nuclei that are small enough for the dynamical timescale to be shorter than the timescale of the activity. Although this has proved possible for some nearby AGN of low to medium luminosity, where observations have directly detected near-nuclear disks and spiral structures, with evidence for gas infall in some cases \citep[see][and references therein]{storchi19}, most quasars are too distant to achieve the required angular resolution. Moreover, their extreme nuclear luminosities can make  near-nuclear structures difficult to study. 

The second major challenge is that the quasar lifetimes are short relative to those of the visible signs of the triggering events on large, kiloparsec scales. For example, in the case of galaxy mergers, the tidal features that characterise such events may remain visible on a $\sim$1\,Gyr timescale \citep[e.g.][]{lotz08} --- a factor 10 -- 1000$\times$ longer than typical quasar lifetimes. Therefore, for every quasar whose host galaxy shows visible tidal features, there are likely to be several galaxies with similar morphological features that do not appear as quasars \citep[see discussion in][]{bess12}. 

Together, these factors mean that it is hard to identify morphological or kinematical features in an {\it individual} quasar host galaxy that uniquely identify the triggering event. Therefore, many studies of quasar triggering have taken the alternative approach of comparing the properties of the {\it population} of quasar host galaxies with those of comparison samples of non-active galaxies. The properties used for comparison include host galaxy morphologies \citep[e.g.][]{heck86,ram11,vill17,ellison19b}, large-scale environments \citep[e.g.][]{serber06,ram13}, star formation rates \citep[e.g.][]{shangguan18,bernhard22} and cold ISM masses \citep[e.g.][]{tad14b,shangguan18,ellison19a,koss21,bernhard22}. 

Given the requirement for high accretion rates that must be sustained over the long timescales of the activity, galaxy mergers are a promising triggering mechanism for quasars, since the associated tidal torques have the potential to deliver large masses of gas to sub-kiloparsec scales on the requisite timescales. This is supported both by hydrodynamical simulations of galaxy mergers \citep{dm05,hopkins08,johansson19,byrne_mamahit22} and by observations of ultra-luminous infrared galaxies \citep[ULIRGs:][]{sanders96}, which show high gas surface densities close to their nuclei and are thought to represent the peaks of major, gas-rich mergers \citep{downes98}. Indeed, it has been proposed that quasars form part of the lifecycles of major, gas-rich mergers: they are triggered close to the peaks of the mergers as the nuclei of the two galaxies coalesce, and then become visible post-coalescence after the enshrouding natal cocoon of gas and dust has been dispersed to reveal the remnant nucleus \citep{sanders88,hopkins08,blecha18}. 

Many investigations of triggering via mergers have used imaging observations to attempt to detect the tidal features (tails, fans, shells etc.) and close companion galaxies expected in merging systems. However, the results appear ambiguous: while some studies find clear evidence for a high rate of morphological disturbance in quasar hosts  \citep{heck86,bahcall97,urrutia08,benn08,ram11,bess12,chiaberge15,urb19,pierce22},  others find no such evidence \citep{dun03,greene09,mechtley16,wyl16,vill17,marian19,zhao19,zhao21}.  Several factors might contribute to this apparent ambiguity, including differences in redshift, surface brightness depth, AGN properties,  spatial resolution, observation wavelength, control selection and classification methodology between the different studies.

In order to overcome these issues, we have undertaken a major imaging study of a sample of nearby AGN ($z < 0.3$) that covers a broad range of optical emission-line luminosity and radio power, but includes a substantial sample of quasars at the high luminosity end. All the AGN were observed to uniform (deep) surface brightness depth in the $r$\,\,band ($3\sigma$ surface brightness depth: $\mu_r \sim 27$\,mag arscec$^{-2}$) using the Wide Field Camera (WFC) on the Isaac Newton Telescope (INT). 
As reported in \citet{pierce22}, the AGN in our sample with high [OIII]$\lambda5007$ emission line luminosities show a substantially enhanced rate of morphological disturbance (by factor $\sim$2 -- 3) compared with the matched control sample galaxies, with the degree of enhancement increasing as a function of [OIII] luminosity (a proxy for AGN bolometric luminosity; e.g. see \citeauthor{heck04} \citeyear{heck04}). This reinforces previous suggestions that the triggering mechanism varies with AGN luminosity \citep{tre12,ellison19b}, although some studies fail to find evidence for such a luminosity dependence \citep[e.g.][]{vill17}.

To further investigate whether galaxy mergers are indeed the dominant triggering mechanism at high AGN luminosities, here we focus on a sample of 48 local ($z < 0.14$) type 2 quasars -- the \href{http://research.iac.es/galeria/cra/qsofeed/}{Quasar Feedback (QSOFEED)} sample \citep{ramos22}.
These are objects that lack the broad permitted lines in their optical spectra that would lead to a type 1 AGN classification, but have strong, high-ionization forbidden emission lines whose luminosities are consistent with those of quasars ($L_{\rm [OIII]} > 10^{35}$ W; \citealt{zak03}). A major advantage of using type 2 quasars for studies of this kind is that they are not affected by the strong nuclear point source emission that may mask the presence of tidal features out to several kiloparsecs in the case of the type 1 objects. Also, at the low redshifts covered by our sample, we avoid major cosmological surface brightness dimming, and have the sensitivity and spatial resolution to detect faint tidal features even in moderate seeing conditions. While preliminary results for 25 objects were included in \citet{pierce22}, here we present observations and results for all 48 type 2 quasars, and draw conclusions about the dominant triggering mechanism for the local type 2 quasar population as a whole. 

The paper is organised as follows. Section 2 describes the sample, observations, reduction and classification methodology. The main imaging results for our type 2 quasar sample are presented in Section 3, where we also compare with those for the matched control sample. Section 4 investigates the reasons for the apparently ambiguous results in the literature. The results are then discussed in the overall context of quasar triggering mechanisms in Section 5, and we present our conclusions in Section 6.  A cosmology with $H_0 = 70$\,km s$^{-1}$ Mpc$^{-1 }$, $\Omega_m = 0.30$ and $\Omega_{\Lambda} =0.70$ is assumed throughout this paper.

\section{Observations and classifications}
\label{sec:obs_and_class}

\begin{figure}
	\includegraphics[width=0.95\linewidth]{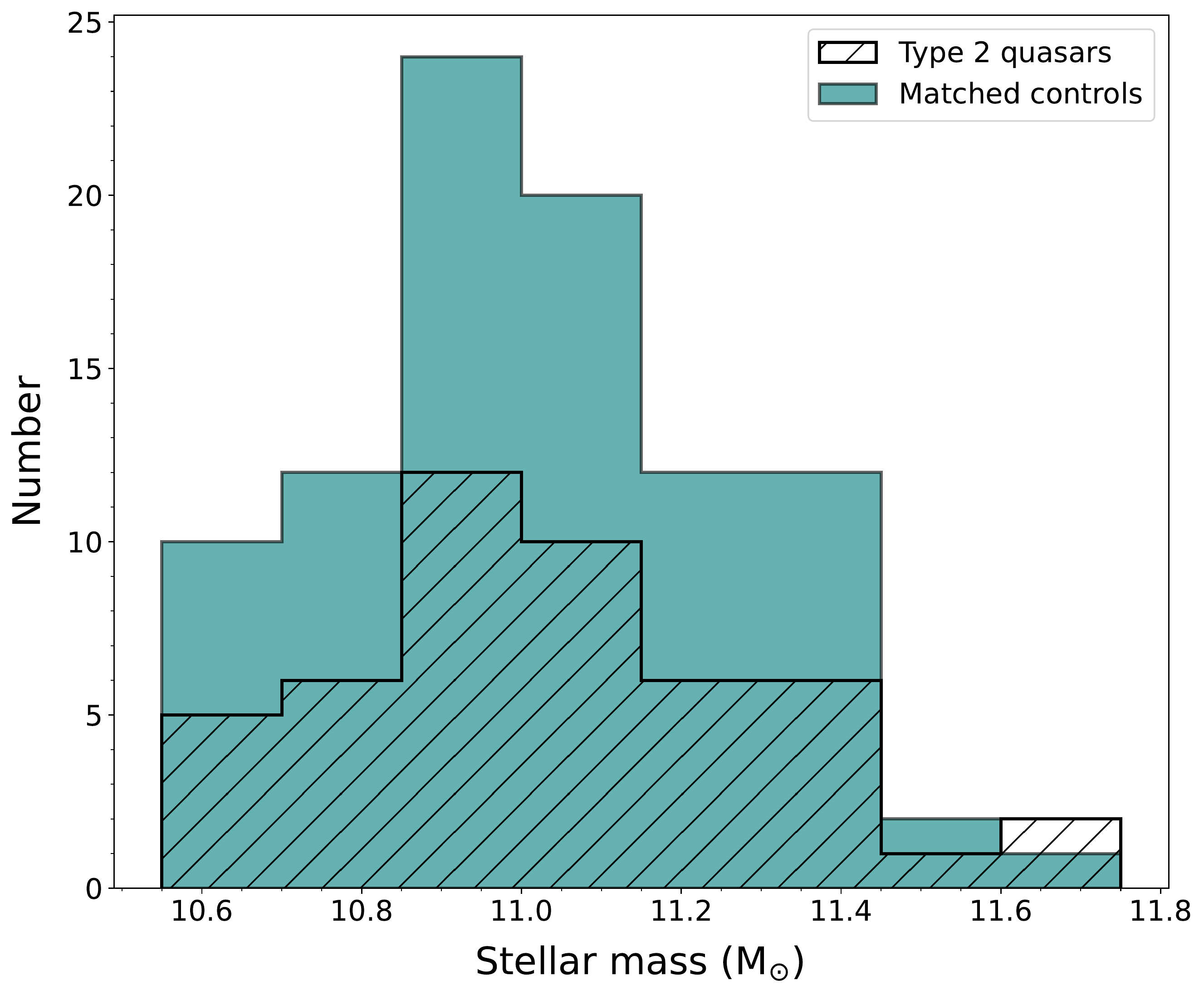}
    \caption{Comparison of the stellar mass distributions of the type 2 quasar host galaxy and control samples.}
    \label{fig:control_comparison}
\end{figure}

\subsection{Sample selection, observations and reductions}
\label{sec:observations}

The sample for this study is that of the QSOFEED project \citep{ramos22}, which is using optical, near-IR and mm-wavelength observations to investigate triggering and feedback in nearby type 2 quasars. It comprises all 48 SDSS-selected type 2 quasars in the compilation of \citet{reyes08} with [OIII]$\lambda$5007 luminosities $L_{\rm [OIII]}>10^{8.5}$\,L$_{\odot}$ ($L_{\rm [OIII]}>10^{35}$\,W), and redshifts $z < 0.14$. Note that, along with [OIII] emission-line luminosity, the criteria for the selection of the type 2 quasar objects in \citet{reyes08} include a lack of the broad permitted lines that would indicate a type 1 AGN, and the necessity that the objects have AGN-like narrow emission-line ratios. Including many of the closest quasar-like AGN in the local universe, the sample is predominantly radio-quiet, with only 4 of the objects having radio luminosities $L_{\rm 1.4GHz} >  10^{24.5}$\,W\,Hz$^{-1}$ that might lead to them being classified as radio-loud AGN. Basic properties of the sample objects are shown in Table \ref{tab:sample}.

\begin{table*}
\centering
\caption{General properties of the type 2 quasar sample. The seventh column gives the 
overall galaxy morphological classification (spiral/disc -- S; elliptical -- E; lenticular -- L; merger -- M; uncertain -- UC), while the penultimate column gives the Zooniverse disturbance classification (UD -- undisturbed; D-Pre -- disturbed, pre-coalescence; D-Post -- disturbed, post-coalescence). Note that, for overall morphological classifications that are uncertain (UC in column 7), the first and second most common votes are shown in brackets (in that order) separated by a comma, whereas an equal vote share is indicated by a slash. The stellar masses listed here have been corrected to the MPA-JHU mass scale \citep[as detailed in][]{pierce22}, and thus differ slightly (by up to $\sim$0.2 dex) from those used in other papers for the QSOFEED sample where this correction has not been made \citep[e.g.][]{ramos22}.}
\label{tab:sample}
\begin{tabular}{ccccccccc}
\hline
\makecell{Full SDSS name}  &\makecell{Abbreviated\\name} &$z$ & $\log_{10}\left(\rm \frac{L_{[OIII]}}{L_{\odot}}\right)$ & $\log_{10}\left(\rm \frac{L_{1.4GHz}}{W\,Hz^{-1}}\right)$ &$\log_{10}\left(\frac{M_*}{M_{\odot}}\right)$ & \makecell{Morph.\\type} &\makecell{Disturb.\\class} &Alternative name  \\
\hline
J005230.59-011548.4 &J0052-0115 &0.1348 &8.58 &23.30 &10.8 &L &UD & \\ 
J023224.24-081140.2 &J0232-0811 &0.1001 &8.60 &23.07 &10.8 &E &UD & \\
J073142.37+392623.7 &J0731+3926 &0.1103 &8.59 &23.03 &11.0 &E &UD & \\
J075940.95+505023.9 &J0759+5050 &0.0544 &8.83 &23.49 &10.6 &E &UD &IRAS\,F07559+5058 \\
J080224.34+464300.7 &J0802+4643 &0.1208 &8.86 &23.68 &11.3 &E &UD & \\
J080252.92+255255.5 &J0802+2552 &0.0811 &8.58 &23.52 &11.1 &M &D-Post & \\
J080523.29+281815.7 &J0805+2818 &0.1284 &8.62 &23.41 &11.4 &UC (S,M) &D-Pre & \\
J081842.35+360409.6 &J0818+3604 &0.0758 &8.53 &22.51 &10.6 &UC (E,L) &D-Pre & \\
J084135.09+010156.3 &J0841+0101 &0.1106 &8.87 &23.32 &11.1 &E &D-Pre & \\
J085810.63+312136.2 &J0858+3121 &0.1387 &8.53 &23.01 &11.1 &E &D-Pre & \\
J091544.18+300922.0 &J0915+3009 &0.1298 &8.78 &23.29 &11.2 &M  &D-Pre & \\
J093952.75+355358.9 &J0939+3553 &0.1366 &8.77 &26.22 &11.0 &E &D-Pre &3C223 \\
J094521.33+173753.2 &J0945+1737 &0.1280 &9.05 &24.28 &10.9 &UC (E,M) &D-Post & \\
J101043.36+061201.4 &J1010+0612 &0.0977 &8.68 &24.34 &11.4 &E &D-Pre & \\
J101536.21+005459.4 &J1015+0054 &0.1202 &8.69 &22.96 &10.9 &UC (E,L) &D-Pre & \\
J101653.82+002857.2 &J1016+0028 &0.1163 &8.63 &23.60 &11.0 &E &D-Pre & \\
J103408.59+600152.2 &J1034+6001 &0.0511 &8.85 &23.06 &11.1 &UC (S,M) &D-Post &Mrk34 \\
J103600.37+013653.5 &J1036+0136 &0.1068 &8.53 &$<$22.45 &11.3 &S &D-Pre & \\
J110012.39+084616.3 &J1100+0846 &0.1004 &9.20 &24.17 &11.4 &S &UD & \\
J113721.36+612001.1 &J1137+6120 &0.1112 &8.64 &25.57 &10.9 &E &UD &4C+61.23 \\
J115245.66+101623.8 &J1152+1016 &0.0699 &8.72 &22.67 &10.8 &UC (E,M) &D-Post &Tololo\,1150+105 \\
J115759.50+370738.2 &J1157+3707 &0.1282 &8.62 &23.37 &11.2 &E &UD & \\
J120041.39+314746.2 &J1200+3147 &0.1156 &9.36 &23.45 &11.0 &E &D-Pre & \\
J121839.40+470627.7 &J1218+4706 &0.0939 &8.58 &22.71 &10.6 &M &D-Post & \\
J122341.47+080651.3 &J1223+0806 &0.1393 &8.81 &$<$22.70 &11.0 &E &UD & \\
J123843.44+092736.6 &J1238+0927 &0.0829 &8.51 &22.30 &11.3 &E &D-Pre & \\
J124136.22+614043.4 &J1241+6140 &0.1353 &8.51 &23.47 &11.4 &M &D-Pre & \\
J124406.61+652925.2 &J1244+6529 &0.1071 &8.52 &23.36 &11.3 &M &D-Pre & \\
J130038.09+545436.8 &J1300+5454 &0.0883 &8.94 &22.69 &10.8 &UC (S,E) &D-Post & \\
J131639.74+445235.0 &J1316+4452 &0.0906 &8.65 &23.07 &11.7 &UC (S/M) &D-Pre & \\
J134733.36+121724.3 &J1347+1217 &0.1204 &8.70 &26.30 &11.7 &M &D-Pre &4C+12.50 \\
J135617.79-023101.5 &J1356-0231 &0.1344 &8.53 &22.93 &11.1 &E &D-Post & \\
J135646.10+102609.0 &J1356+1026 &0.1232 &9.21 &24.38 &11.3 &M &D-Post & \\
J140541.21+402632.6 &J1405+4026 &0.0806 &8.78 &23.41 &10.8 &E &D-Post &IRAS\,F14036+4040 \\
J143029.88+133912.0 &J1430+1339 &0.0851 &9.08 &23.67 &11.1 &UC (E/M) &D-Post &`The Teacup' \\
J143607.21+492858.6 &J1436+4928 &0.1280 &8.61 &23.59 &11.0 &E &D-Pre & \\
J143737.85+301101.1 &J1437+3011 &0.0922 &8.82 &24.15 &11.2 &UC (M,E) &D-Post &IRAS\,F14354+3024 \\
J144038.10+533015.9 &J1440+5330 &0.0370 &8.94 &23.28 &10.6 &E &D-Pre &Mrk477 \\
J145519.41+322601.8 &J1455+3226 &0.0873 &8.64 &22.73 &10.6 &UC (S,E) &D-Post & \\
J150904.22+043441.8 &J1509+0434 &0.1114 &8.56 &23.82 &10.9 &S &UD & \\
J151709.20+335324.7 &J1517+3353 &0.1353 &8.91 &24.75 &11.5 &E &UD &IRAS\,F15151+3404 \\
J153338.03+355708.1 &J1533+3557 &0.1286 &8.56 &$<$22.62 &10.9 &UC (E/L) &UD & \\
J154832.37-010811.8 &J1548-0108 &0.1215 &8.52 &$<$22.57 &10.8  &E &UD & \\
J155829.36+351328.6 &J1558+3513 &0.1215 &8.77 &23.19 &10.9 &E &UD & \\
J162436.40+334406.7 &J1624+3344 &0.1224 &8.56 &23.11 &11.0 &UC (E/S) &D-Pre & \\
J165315.05+234942.9 &J1653+2349 &0.1034 &9.00 &23.38 &11.0 &E &UD &`The Beetle' \\
J171350.32+572954.9 &J1713+5729 &0.1128 &8.99 &23.42 &11.1 &E &UD & \\
J215425.74+113129.4 &J2154+1131 &0.1092 &8.54 &23.25 &10.9 &E &UD & \\
		\hline
	\end{tabular}
\end{table*}

The sample was imaged with the Wide Field Camera (WFC)  on the Isaac Newton Telescope (INT) at the Observatorio del Roque de los Muchachos, La Palma. A particular advantage of the INT/WFC is its wide field of view (34'$\times$34'). This has allowed us to leverage a large control sample of non-active galaxies in the fields of our wider AGN sample using images of this sample taken over several observing runs in the last decade (described here and in \citeauthor{pierce22} \citeyear{pierce22}; total area imaged $>$50\,deg$^{2}$). The fact that the control galaxies were observed simultaneously with the AGN hosts in the same images ensures that the average image quality and depth are the same for both groups.  

\begin{figure*}
	\includegraphics[width=17.7cm]{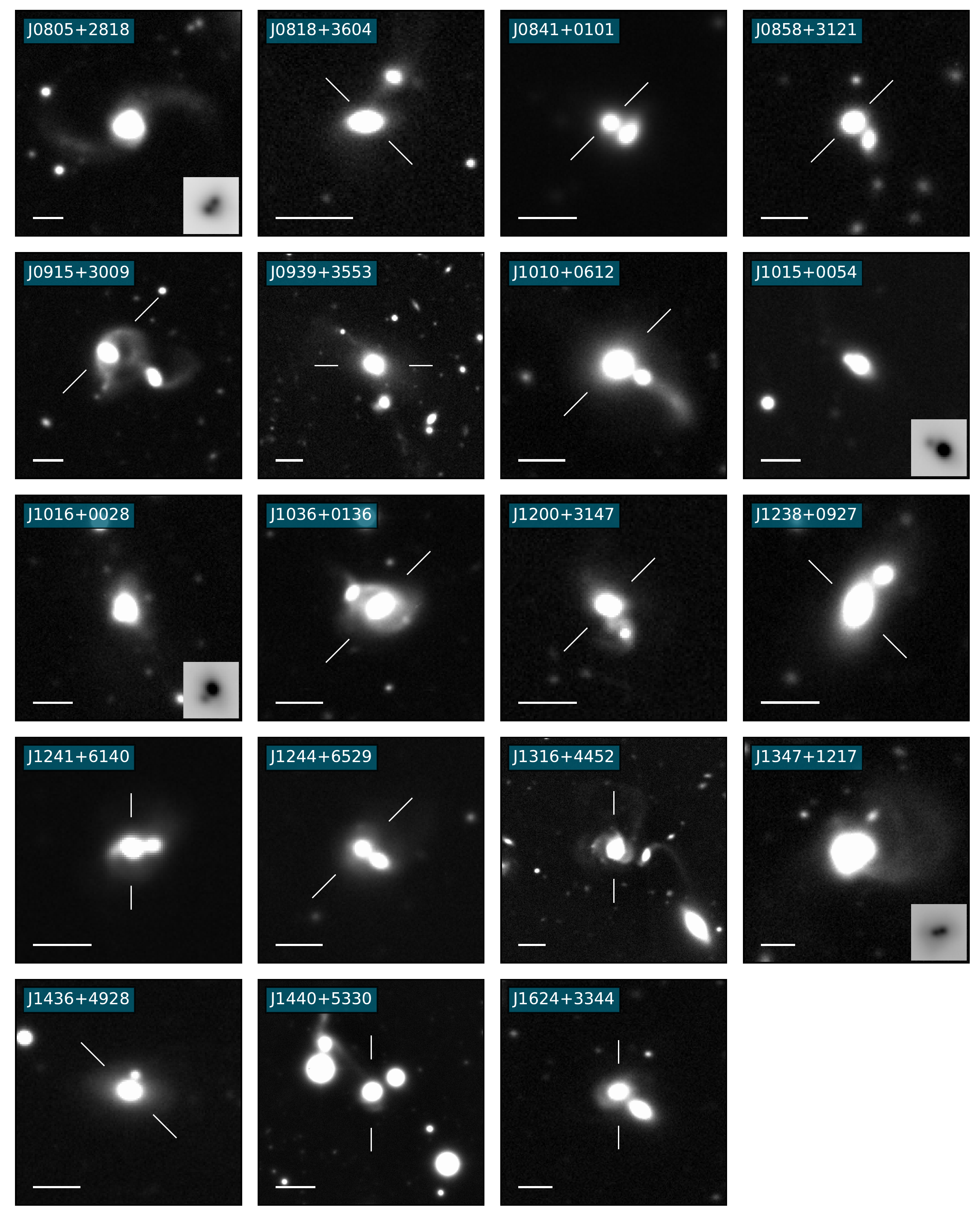}
    \caption{Thumbnail images of the 19 type 2 quasar objects classified as pre-coalescence systems. In each case north is to the top and east to the left, and the bar in the bottom left-hand corner indicates a distance of 20\,kpc at the redshift of the source. The type 2 quasar object is generally situated close to the centre of the image, but in cases where there might be ambiguity due to close companion galaxies, the type 2 quasar is identified by line segments. Insets show the close double nuclei detected in some cases. }
    \label{fig:pre_coalescence}
\end{figure*}

The 48 type 2 quasar objects were observed with the INT/WFC in two separate observing runs: 25 with right ascensions in the range 00h 00m $<$ RA $<$ 12h 30m in January 2020 \citep[see][]{pierce22}, and 23 with right ascensions in the range 12h 30m $<$ RA $<$ 22h 00m in May 2021. Full details of the observing strategy and reduction of the imaging data are given in \citet{pierce22}, but are summarised briefly here. The images were taken with the WFC Sloan $r$-band filter, using four separate dithered 700\,s exposures (separated by 30 arcsec to cover the CCD chip gaps) to achieve a typical total exposure time of 2800\,s for each object. The CCD images were then bias-subtracted, flat fielded and combined using \textsc{THELI} \citep{schirmer13}, which also performed astrometric and photometric calibration using catalogued Pan-STARRS1 \citep[][]{cham16} stars detected in the image fields. The pixel scale of the reduced images is 0.333 arcsec pixel$^{-1}$, and based on measurements of the full-widths at half maximum ($FWHM$) of stars in the reduced images, the seeing conditions were moderate for the observations: $0.9 < FWHM < 2.0$\,arcsec for the January 2020 run, and $1.3 < FWHM < 2.4$\,arcsec for the May 2021 run; seeing estimates for the observations of individual objects are presented in Table \ref{tab:obs_sb} in Appendix~\ref{sec:appendix}. In line with our previous INT/WFC observations with the same exposure times \citep{pierce19,pierce22}, we achieved a typical surface brightness depth of $\mu_r = 27.0$\,mag arcsec$^{-2}$ across both runs.\footnote{Consistent with \citet{pierce22}, and the surface brightnesses of the faintest features measured in our images, this is a 3$\sigma$ surface brightness depth estimated from the standard deviation ($\sigma$) of the flux measurements obtained using multiple circular sky apertures of 1\,arcsec in radius (area: 3.14\,arcsec$^2$).}

Although the primary analysis in this paper is based on the INT/WFC camera images described above, in order to investigate the effect that seeing conditions may have on our results, a subset of 6 objects in our sample (J1436+4928, J1455+3226, J1517+3353, J1533+3557, J1548-0108, J1558+3513) was observed in June 2021 in better seeing conditions ($0.9 < FWHM < 1.20$ arcsec) using an $r$-band filter with the PF-QHY camera on the William Herschel Telescope (WHT), on La Palma. The PF-QHY camera contains a back-illuminated CMOS detector (Sony IMX455) with a field of view of 10.7$\times$7.1 arcminutes and 0.067 arcsecond pixels. Using a four point dither pattern with 450\,s exposures at each position, the total exposure time per object was 1800\,s. During the reduction of the images with \textsc{THELI}, the data were resampled to a pixel scale of 0.333 arcsec pixel$^{-1}$ to allow direct comparison with the INT/WFC observations. A similar surface brightness depth ($\mu_r = 27.1$\,mag arcsec$^{-2}$) was achieved for the WHT/PF-QHY  observations as for the INT/WFC observations. Thumbnails of the reduced images are shown in Figure  \ref{fig:wht} in Appendix~\ref{sec:appendix}. We emphasise that these
images were not used for the main morphological classification described below, but they nonetheless provide useful supplementary information (see Sections \ref{sec:results_rates} and \ref{sec:results_class}).

\begin{figure*}
	\includegraphics[width=17.6cm]{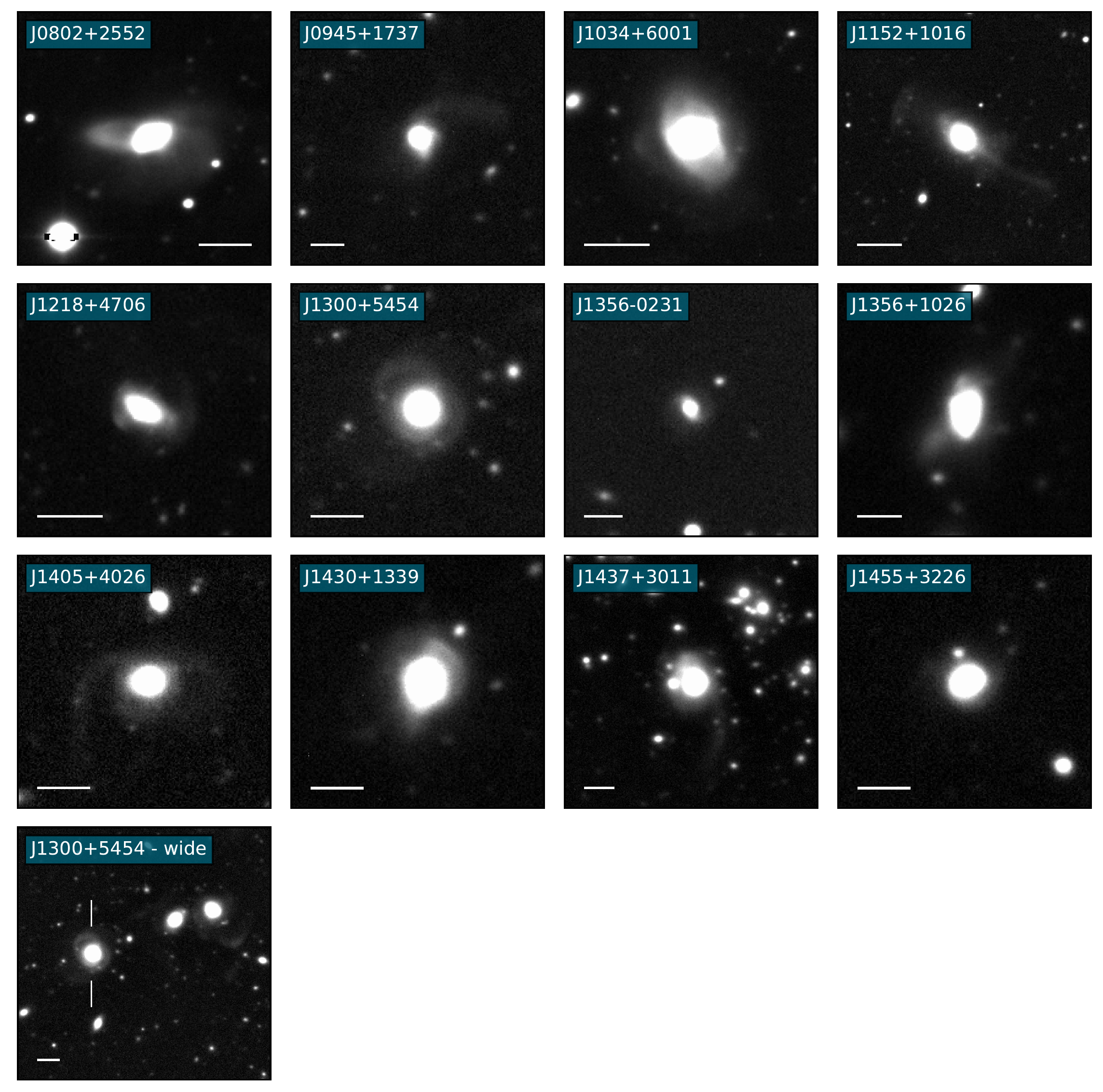}
    \caption{Thumbnail images of the 12 type 2 quasar objects classified as post-coalescence systems. In each case north is to the top and east to the left, and the bar in the bottom left- or right-hand corner indicates a distance of 20\,kpc at the redshift of the source. The lowest-left panel shows the wider field surrounding J1300+5454 (indicated by line segments), demonstrating that it is part of an interacting group of galaxies. }
    \label{fig:post_coalescence}
\end{figure*}

As described in detail in \citet{pierce22}, stellar masses for the type 2 quasar objects were derived using their 2MASS $K_s$-band magnitudes and corrected to the same mass scale as the MPA-JHU value-added catalogue for SDSS DR7. The non-active control galaxies were selected from the MPA-JHU catalogue based on matching in position (to be in one of our previously observed INT/WFC fields), redshift (tolerance: $\pm$0.01 in $z$) and stellar mass (overlap of the one-sigma uncertainties of the quasar host and comparison galaxy mass measurements). Galaxies listed in the catalogue as AGN or broad-line (but not star-forming) objects were not considered for the matching. Typically 2 -- 5 stellar mass- and redshift-matched control galaxies were selected for each type 2 quasar host galaxy, resulting in a total of 125 unique control galaxies after controls selected for more than one type 2 quasar host were accounted for. We note that none of the control galaxies are close enough to be interacting with the quasar 2 host galaxies.

When comparing the stellar mass distribution of this initial control selection to that of the type 2 quasar sample, it was found that the control sample showed a significantly higher proportion of lower-stellar-mass galaxies compared to the type 2 quasar sample, a consequence of the fact that more control galaxies were selected on average for the lower, than the higher, stellar mass type 2 quasar objects. This is potentially a source of concern, because the results of \citet{pierce22} demonstrated that there is a mild increase in the proportion of control galaxies showing evidence for morphological disturbance with stellar mass: from $\sim$20 per cent at $M_* = 10^{10.5}$\,M$_{\odot}$ to $\sim$30 per cent at $M_* = 10^{11.5}$\,M$_{\odot}$\footnote{Note that no evidence was found in \citet{pierce22} for a change in disturbance rate with redshift over the redshift range of our type 2 quasar sample. Therefore, in terms of selecting the control sample, we have prioritised matching the mass rather than the redshift distribution.}. In order to counteract any potential biases that this might introduce, we therefore added a further stellar-mass constraint on our control sample. We first divided the stellar mass distribution of the type 2 quasar hosts into 8 equal logarithmic bins in the mass range $10.55 < \log(M_*/M_{\odot}) < 11.75$. Then, where possible, we randomly selected exactly twice as many control galaxies as type 2 quasars in each bin.
The resulting stellar mass distribution of the 93 objects in a stellar-mass-constrained control sample selected in this way is compared with that of the type 2 quasar objects in Figure \ref{fig:control_comparison}.
Note that each stellar mass bin has two randomly-selected control galaxies per type 2 quasar, apart from the highest mass bin, which has only 0.5 controls per type 2 quasar. Since this highest stellar mass bin contains only two type 2 quasar objects (4 per cent of the full sample), the relative dearth of controls in this bin does not significantly affect our overall results or conclusions. In the following, all the statistics relating to the comparison between the type 2 quasar hosts and control galaxies are based on the average of 100 random selections of the control galaxy stellar mass distribution.

\begin{figure*}
	\includegraphics[width=17.5cm]{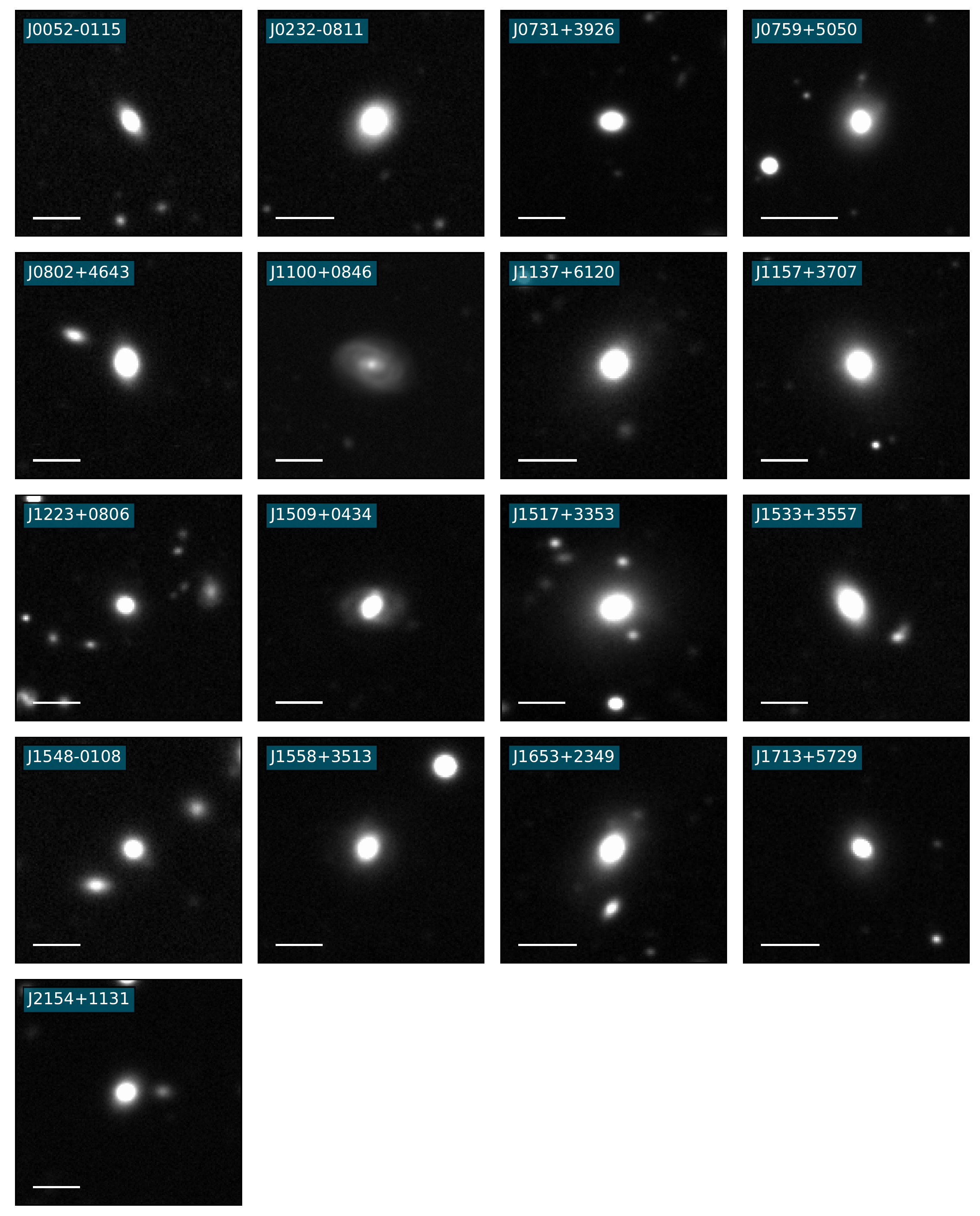}
    \caption{Thumbnail images of the 17 type 2 quasar objects classified as undisturbed systems. In each case north is to the top and east to the left, and the bar in the bottom left-hand corner indicates a distance of 20\,kpc at the redshift of the source.  }
    \label{fig:undisturbed}
\end{figure*}

\subsection{Classification and surface brightness}
\label{sec:classification}

The type 2 quasar hosts and their controls were classified by eight expert classifiers (all the authors of this paper, except PB, GS and JH) via a web-based interface run using the Zooniverse platform (\citealt{lin08}, \citealt{lin11}; see \citealt{pierce22} for full details). The classifiers were presented with two images scaled to have standard area 200\,kpc $\times$ 200\,kpc at the redshift of each object, one at high and one at low contrast; note that the classifiers could not manually change the contrast levels within the Zooniverse platform. In order to prevent biases, the images of the type 2 quasar hosts and controls were classified together and presented in random order, such that in each case the classifiers did not know whether they were classifying a type 2 quasar host or a control galaxy.

For each object, the classifiers were first asked whether the object shows morphological evidence for galaxy interactions, with possible answers of `Yes', `No' or `Not classifiable' (e.g. because of image defects). Then, when the answer to this first question was `Yes', the classifiers were asked to make a more detailed classification by selecting one or more of the following morphological categories: tail (T); fan (F); shell (S); amorphous halo (A); irregular (I); bridge to companion (B); multiple nuclei within a projected distance  of 10\,kpc (MN); and tidally-interacting companion (TIC). Note that the latter three categories (B, MN, TIC) indicate a merger or galaxy interaction in a pre-coalescence phase, whereas the other features (T, F, S, A, I)  could be present pre- or post-coalescence. Therefore, we consider objects whose features include those classified as B, MN or TIC to be pre-coalescence systems, and those showing {\it only} T, F, S, A or I features to be post-coalescence systems \citep[similar to][]{ram11,ram12,bess12}.

We emphasise that, to be classified as a pre-coalescence system, objects lacking multiple nuclei within 10\,kpc (type MN) had to show a clear bridge to a companion (type B) or evidence for tidal features/distortion in the host or companion associated with their interaction (type TIC). The mere presence of a companion close to the host but outside 10\,kpc is not regarded as sufficient evidence for the object to be classified as a disturbed pre-coalescence system, unless there is independent evidence (types B and TIC) for interaction. Thus, our classification scheme is conservative. For example, J0802+4643, J1533+3557 and J2154+1131 (see Figure \ref{fig:undisturbed}) are not classified as disturbed, despite the fact that they are all potentially interacting with companion galaxies within $\sim40$\,kpc.

On the basis of these classifications, objects were classified as disturbed (i.e. showing evidence for galaxy interactions) or undisturbed, and the disturbed objects were further sub-divided into pre- or post-coalescence systems, based on a simple majority of the selections made by the classifiers who had classified the galaxy as disturbed. Thumbnail images for objects classified as pre-coalescence, post-coalescence and undisturbed are shown in Figures \ref{fig:pre_coalescence}, \ref{fig:post_coalescence} and \ref{fig:undisturbed} respectively, and the results of the classifications are presented in Table \ref{tab:sample}. 

Note that the distinction between pre- and post-coalescence is not always clear-cut. For example, J1300+5454 (Figure \ref{fig:post_coalescence}) is situated in a galaxy group in which multiple galaxy interactions are taking place. It has been classified as a post-coalescence system using the Zooniverse interface, presumably on the basis that it shows tidal features at modest radial distances ($<$50\,kpc), but no nearby companion or secondary nucleus is detected. However, on larger scales it shows evidence for tidal interaction with two galaxies of similar brightness $\sim$100 -- 150\,kpc to the NW (see lowest-left panel in Figure \ref{fig:post_coalescence}). The latter interaction might have induced the tidal features detected at smaller radial distances in the quasar host and led to the triggering of the quasar activity. If so, a pre-coalescence classification would be more appropriate.

As well as the detailed morphology, the classifiers were  asked to indicate which of the following labels best described the  overall morphological type of the galaxy: spiral/disc (S); elliptical (E); lenticular (L); merger (M: too disturbed to classify as one of the standard galaxy types). As for the detailed classifications, the final overall morphological type was determined based on a simple majority of the Zooniverse votes. Objects for which no majority was obtained for any particular type were classified as uncertain (UC). The results are shown in Table \ref{tab:sample}.

Following the completion of the Zooniverse classification, we measured the surface brightnesses of any tidal features detected (T, F, S, A, I, B and TIC categories above), again using measurement apertures of 1 arcsec in radius. In some cases it was necessary to subtract the diffuse background light of the galaxy, in order to measure the surface brightness of the feature accurately. In such cases, we used one of the following three approaches depending on the feature and its position in the galaxy: background apertures on either side of the feature in the azimuthal direction but at the same radial distance from the galaxy nucleus as the features; background apertures at the same radial distance on the opposite side of the galaxy; or background apertures along the same position angle but different radial distances, with one closer to the nucleus and one further from the nucleus than the feature. 

The background-subtracted $r$-band surface brightnesses were corrected to the rest frame by applying Galactic extinction, surface brightness dimming and K\,corrections, as described in \citet{ram12}. Finally, to allow comparisons with other studies, the surface brightnesses were converted to the $V$-band using the $V-r$ colour for an Sbc galaxy\footnote{Without further photometric or spectroscopic information, the colours and SEDs of the tidal features are unknown. However, assuming an Sbc colour can be seen as a reasonable compromise between red elliptical and bluer starburst colours.} from \citet{fuk96}; an Sbc colour was also assumed when making the K correction. Note that, although for some objects we measured the surface brightnesses of several features, in the following analysis we will only consider the highest surface brightness feature in each object. The detection of such features would be most likely to lead to an object being classified as disturbed, and by concentrating on them we avoid biasing our surface brightness comparisons to spectacular objects that show several features.
The distribution of brightest-feature surface brightness is shown for the 31 disturbed objects in the type 2 quasar sample in Figure~\ref{fig:sb_comp}, and the surface brightnesses of all the tidal features measured in individual objects are presented in Table \ref{tab:obs_sb} in Appendix~\ref{sec:appendix}.

\section{Results}
\label{sec:results}

\subsection{The rate of morphological disturbance}
\label{sec:results_rates}

A clear majority of the 48 type 2 quasar objects in our sample -- 65$^{+6}_{-7}$ per cent\footnote{Throughout this paper, uncertainty estimates for proportions have been obtained by following the method of \cite{cameron11}. This involves calculating binomial population proportion uncertainties using a Bayesian approach. Note that these uncertainties are based purely on the proportions, and do not take into account the uncertainties caused by the subjective nature of the classifications and variations amongst the classifiers.}  -- show signs of morphological disturbance that are consistent with them being involved in galaxy interactions. We obtain similar results if we consider the objects from the two observing runs separately (64$^{+8}_{-10}$ per cent and 65$^{+8}_{-11}$ per cent for the objects observed in January 2020 and May 2021 respectively), despite the fact that the seeing was significantly poorer for the second run. In contrast, the mass- and redshift-matched control galaxies show a much lower rate of morphological disturbance: 22$^{+5}_{-4}$ per cent. 
A two-proportion Z-test shows that the difference between the type 2 quasar hosts and the controls is significant at the 5$\sigma$ level. Overall, the hosts of the type 2 quasars show a disturbance rate that is enhanced by a factor 3.0$_{-0.8}^{+0.5}$ relative to that of control galaxies of similar stellar mass and redshift.

\begin{figure}
	\includegraphics[width=8.5cm]{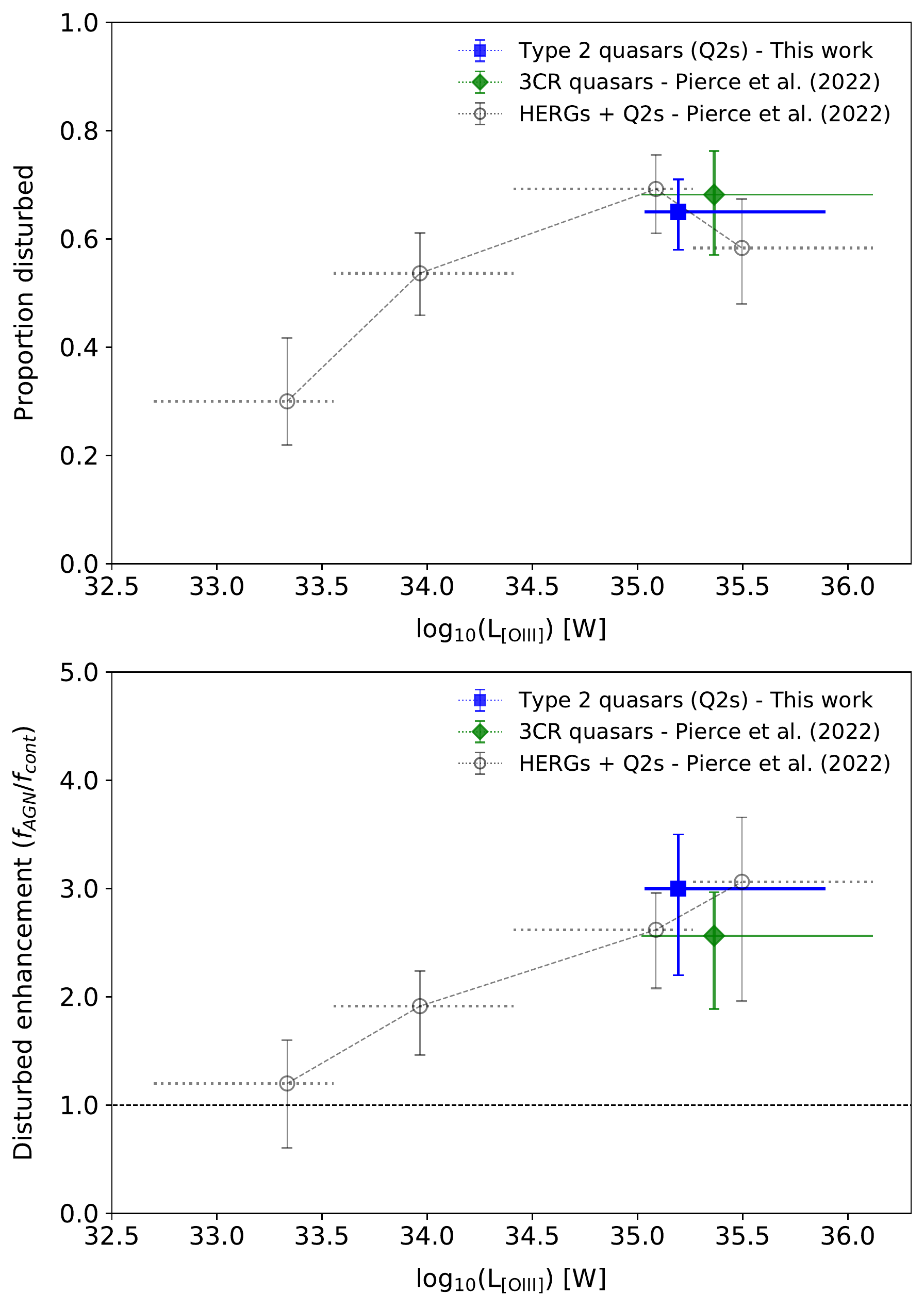}
    \caption{Fraction of disturbed host galaxies (top) and the disturbance enhancement factor relative to controls (bottom) plotted against [OIII] emission-line luminosity ($\rm L_{[OIII]}$), a proxy for the AGN bolometric luminosity.
    The point for the type 2 quasar hosts (blue square) is consistent with the trend of increasing disturbance fraction and enhancement factor with $\rm L_{[OIII]}$ found by \citet{pierce22} for a large sample of high excitation radio galaxies (HERG) and type 2 quasars (grey circles). A point for the 3CR radio quasars ($z < 0.3$) is also shown for comparison (green diamond). The horizontal lines show the [OIII] luminosity range covered by each sample or luminosity bin. Note that the results shown for the type 2 quasars from this study and the 3CR radio quasars from \citet[][]{pierce22} are not entirely independent of points for the HERG and type 2 quasar sample of \citet{pierce22}, which includes the results for many of these objects.}
    \label{fig:merger_frac}
\end{figure}

To put these results into context, we compare them with those of \citet{pierce22} in Figure \ref{fig:merger_frac}. Clearly, the type 2 quasars are consistent with the trend of increasing host galaxy disturbance rate and enhancement factor with [OIII] emission-line luminosity.

These results are likely to represent  lower limits on the true rates of morphological disturbance in both the type 2 quasar hosts and their controls. As discussed in \citet{pierce22}, the fact that the classifiers had limited ability to vary the contrast levels and sizes of the images in Zooniverse means that some morphological signs of disturbance are likely to have been missed. Indeed, among the objects classified in Zooniverse as undisturbed, four objects (J0759+5050, J1517+3353, J1558+3513 and J1653+2349) show clear evidence for tidal features on a closer inspection that allows full manipulation of the contrast levels. Moreover, the modest seeing conditions for the observations are likely to have acted against the detection of finer morphological features and some close double nuclei.  This is demonstrated by our better-seeing WHT/PF-QHY observations (see Figure \ref{fig:wht}), which confirm the large-scale tidal features in J1558+3513, and reveal second nuclei within 10\,kpc in the cases of J1517+3353 and J1548-0108\footnote{This object was classified in Zooniverse as undisturbed, although a faint extension to the SW of main body of the host galaxy is clearly visible in our INT/WFC observations (see Figure \ref{fig:undisturbed}). However, our WHT/PF-QHY image (see Figure \ref{fig:wht}) shows that this faint extension comprises a distinct second nucleus.} that were missed in our INT/WFC observations.

\subsection{Detailed classifications}
\label{sec:results_class}

A striking aspect of the detailed morphologies of the type 2 quasar host galaxies is the wide diversity of structures present (see Figures \ref{fig:pre_coalescence} and \ref{fig:post_coalescence}), with a variety of pre- and post-coalescence features detected. This provides further evidence that luminous, quasar-like AGN are not all triggered at a single evolutionary stage of a particular type of merger \citep[see also][]{ram11,bess12}. However, among the objects classified as morphologically disturbed, we do find a preference for pre- over post-coalescence systems, with 61$^{+8}_{-9}$ per cent of disturbed type 2 quasar hosts classified as pre-coalescence.

Although imaging observations alone cannot provide a precise indication of whether the pre-coalescence systems represent major or minor mergers (usually defined as below or above a mass ratio of 1:4, respectively), based on the fluxes of the residual bulges of the two interacting galaxies, 9 objects (47 per cent of the pre-coalescence systems) are strong candidates for major merger systems: J0805+2818; J0841+0101; J0858+3121; J0915+3009; J1241+6140; J1244+6529; J1347+1217; J1440+5330; and J1624+3344. Indeed,
J1347+1217 (also known as 4C+12.50 or PKS1345+12) is a well-known ULIRG.  

Note that, we have labelled objects with B, MN or TIC classifications as pre-coalescence systems. However, since we lack detailed information on the stellar and gas kinematics, we cannot absolutely guarantee that these systems will eventually coalesce. Therefore, it is more accurate to describe the disturbed systems in our sample as a whole as having undergone galaxy interactions, since this encompasses merged systems (i.e. post-coalescence), systems that will eventually merge, and those that have undergone a close encounter but will not merge.

\begin{table*}
\begin{minipage}{175mm}
\caption{Comparison of results from various imaging-based studies of quasar host galaxies. The second column indicates the optical classes of the quasars studied and, where known, whether they are radio-loud (RL) or radio-quiet (RQ). Where possible, the uncertainties on the disturbance fractions were obtained using the method of \citet{cameron11}, for consistency. The fifth column gives the quoted surface brightness depth (where an estimate is given in the paper), with the filter(s) this was measured for shown in brackets.}
\label{tab:comparison}
\begin{tabular}{lllllll}
\hline
Study &Type &z &N  &\makecell{SB Depth\\(mag arcsec$^{-2}$)}   &\makecell{Disturbance\\fraction} &Notes \\  
\hline 
{\bf HST} & & & & & &\\
Dunlop et al. (2003) &Type 1 (RL+RQ) &0.1--0.25 &23  &23.8 (R) &$\sim$43 per cent\footnote{Based on examination of the psf-subtracted images in \citet{mclure99} and \citet{dun03}. Note that \citet{dun03}  do not provide an estimate of the overall rate of morphological disturbance for the quasars in their full sample.} & Excluding radio galaxies \\
Bennert et al. (2008) &Type 1 (RL+RQ) &0.15--0.21 &5  &-- &80 per cent & \\ 
Urrutia et al. (2008) &Type 1  &0.41--0.95 & 15  &-- &85 per cent &Red quasars \\ 
Wylezalek et al. (2016) &Type 2 &0.2--0.6 &20  &-- &45$^{+11}_{-10}$ per cent & \\
Villforth et al. (2017) &Type 1  &0.5--0.7 &20  &-- &$\sim$25 per cent & \\
Mechtley et al. (2016) &Type 1  &1.9--2.1 &9  &-- &39$\pm$11 per cent &Quasars with most massive SMBHs \\
Urbano-Mayorgas et al. (2019) &Type 2 &0.3--0.4 &41   &-- &34$^{+6}_{-9}$ per cent &Based on HST snapshot images \\
Zhao et al. (2019) &Type 2 &0.04--0.4 &29  &25 (B,I) &34$^{+10}_{-8}$ per cent & \\
Zhao et al. (2021) &Type 1  &$<$0.5 &35  &25 (B,I) &$\sim$20 per cent & \\
Marian et al. (2019) &Type 1 &1.81--2.15 &21  &--  &24$\pm$9 per cent & \\
\vspace{-0.3cm} \\
\bf{Ground-based} & & & & & &\\
Greene et al. (2009) &Type 2  &0.1--0.45 &15  &24--25 (g,r,i) &27$^{+14}_{-8}$ per cent & \\
Ramos Almeida et al. (2011) &Type 1 \& 2 (RL) &0.05--0.7 &26  &$\sim$27 (r) &96$^{+1}_{-8}$ per cent &2Jy objects with $L_{\rm [OIII]} >10^{35}$\,W \\
Bessiere et al. (2012) &Type 2 (mainly RQ) &0.3--0.41 &20  &$\sim$27 (r) &75$^{+7}_{-12}$ per cent & \\
Marian et al. (2020) &Type 1 &0.12--0.19 &17  &$\sim$23.4 (B,V) &41$^{+12}_{-10}$ per cent &High Eddington ratio objects \\
Pierce et al. (2022) &Type 1 \& 2 (RL) &0.05--0.3 &21  &27.0 (r) &67$^{+8}_{-11}$ per cent &3CR objects with $L_{\rm [OIII]}>10^{35}$\,W\\
This work &Type 2 (mainly RQ) &0.04--0.14 &48  &27.0 (r) &65$^{+6}_{-7}$ per cent  & \\
		\hline
	\end{tabular}
\end{minipage}
\end{table*}

\subsection{Overall morphological type}
\label{sec:morph}

Based on the results presented in Table \ref{tab:sample}, half of the host galaxies are classified as ellipticals (E: 50 $\pm$ 7 per cent). However, objects with a majority spiral/disc (S: 6$^{+5}_{-2}$ per cent) or lenticular (L: 2$^{+4}_{-1}$ per cent) vote are rare.     
The remainder are classified as either mergers that are too disturbed to certainly assign
one of the E, S or L morphological types (M: 15$^{+7}_{-4}$ per cent) or uncertain (UC: 27$^{+7}_{-5}$ per cent). Of the latter UC type, in 54 per cent of cases (15 per cent of full type 2 quasar sample) the merger (M) category is the 1st or 2nd most common vote. This indicates that it is sometimes challenging to determine an overall galaxy classification for host galaxies that are involved in mergers. Moreover, the relatively modest spatial resolution of our INT/WFC observations may lead to compact spiral structures being missed. This point is emphasised by the cases of J1455+3226 and J1558+3513, which were classified as UC and E respectively on the basis of Zooniverse classification of their INT/WFC images, but show evidence for inner spiral structures in WHT/PF-QHY images taken in better seeing conditions (see Figure \ref{fig:wht}). If we include all the uncertain objects for which the S or L classification is first or second in terms of votes, the proportion of type 2 quasar hosts that have such morphologies rises to 21 per cent; however, this is likely to be an upper limit.

For comparison, the average morphological type classifications for the stellar-mass-restricted control galaxy selections are: 53 $\pm$ 5 per cent ellipticals, 33$^{+5}_{-4}$ per cent spirals, 4$^{+3}_{-1}$ per cent lenticulars, 1$^{+2}$ per cent mergers, 9$^{+4}_{-2}$ per cent uncertain. Although there is a higher proportion of spirals in the control sample, we do not expect this to affect the comparison between control and quasar 2 disturbance fractions, because in \citet{pierce22} we demonstrated that late- and early-type non-active control galaxies have similar disturbance fractions.

Our results are consistent with other studies that find a high proportion of early-type host galaxies for quasars \citep[e.g.][]{dun03,urb19}. In particular,  perhaps the most directly comparable  study is that of \citet{urb19}, who visually classified the morphologies of 41 type 2 objects with quasar-like luminosities ($L_{\odot} > 10^{8.3}$) in HST snapshot images, and found 
66$^{+5}_{-10}$ per cent to be hosted by ellipticals and 12$^{+7}_{-4}$ per cent by spirals/disks -- these results are entirely consistent with proportions found for
our sample, given the uncertainties.\footnote{Note that parametric modelling of the surface brightness profiles of quasar hosts has the potential to reveal faint disk structures that are not clearly apparent on visual inspection. The detection of such disks could increase the proportion of objects in our sample that show some evidence
for the presence of disk components. However, the limited spatial resolution of our INT data precludes 
such analysis.}

\section{Comparison with disturbance rates in the literature}
\label{sec:comparison}

As noted in the Introduction, there is considerable ambiguity in the literature surrounding the dominant triggering mechanism for quasars. Table \ref{tab:comparison} summarises the results of various quasar imaging studies from the last 20 years. 

It is challenging to compare the different studies because they are based on data taken with different telescope/instrument/filter combinations to different effective surface brightness depths. They also cover a range of redshifts, types of quasars (type 1, type 2, radio-loud, radio-quiet) and classification methodologies. However, one clear pattern emerges: most of the HST-based studies give a relatively low rate of morphological disturbance (20 to 45 per cent), whereas the majority of the ground-based studies -- including our current study of nearby type 2 quasars -- give a high rate of disturbance (65 to 96 per cent). The exceptions are the \citet{benn08} and \citet{urrutia08}\footnote{\citet{urrutia08} study a sample of quasars selected on the basis of red optical/near-IR colours. This selection may bias the sample to objects that have recently undergone a merger.}   studies, which give a much higher rates of disturbance than the other HST studies, and the \citet{greene09} and \citet{marian20} studies, which give lower rates of disturbance than other ground-based studies. Significantly, the \citet{benn08} study is much deeper than the other HST studies (5 orbits compared with $\le$1 orbit), while the \citet{greene09} and \citet{marian20} studies have significantly lower quoted surface brightness depths than the other ground-based studies. This suggests that the surface brightness depth of the observations is a key factor -- a point highlighted by the \citet{benn08} study, which detected low-surface-brightness tidal features that were not detected in the shallower HST observations by \citet{dun03} of the same objects (see also \citealt{ellison19b} and \citealt{wilkinson22} for discussion of ground- and simulation-based results for different surface brightness depths).

Along with the observational surface brightness depth, $(1+z)^4$ cosmological surface brightness dimming will also have a major impact on the detectability of tidal features. This dimming amounts to 2.0\,mag by $z=0.6$ and 4.8\,mag by $z=2$, substantially reducing the sensitivity of studies of high redshift quasars to detecting the signs of morphological disturbance.

\begin{figure*}
	\includegraphics[width=0.85\linewidth]{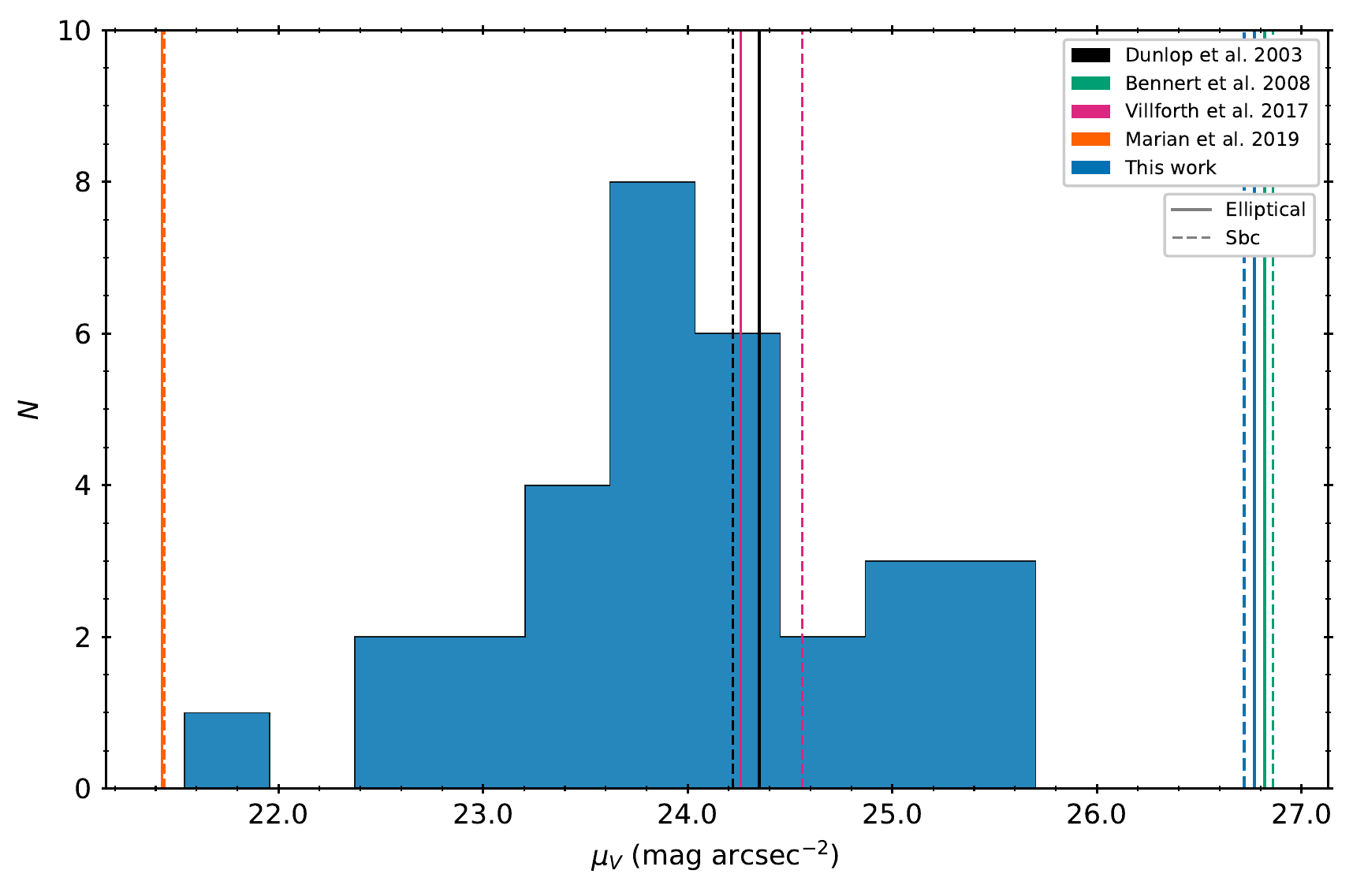}
    \caption{Histogram showing the distribution of surface brightness estimated for the tidal features in the 31 type 2 quasar objects classified as disturbed. Only the highest surface brightness feature detected in each object is included. The vertical lines show the estimated limiting surface brightness depths for various studies reported in the literature, with two lines shown for each object: one calculated assuming the colour of an elliptical galaxy (solid), and the other calculated assuming the colour of an Sbc galaxy (dashed) -- see Appendix~\ref{sec:appendix} for details. The limiting surface brightness depths for the current study (blue lines) have been estimated by assuming a redshift of $z=0.11$, which is typical of our type 2 quasar sample. Note that, in the case of the \citet{marian19} study, the lines
    for the two assumed colours are coincident. For ease of comparison, all surface brightness estimates have been transformed to the rest-frame V-band, as described in Section \ref{sec:classification}.}
    \label{fig:sb_comp}
\end{figure*}

We have explored the combined effects of observational surface-brightness depth and dimming by estimating the effective depths of some key studies in the literature. It is difficult to compare the depths using the quoted values in the studies themselves because, where values for the limiting surface brightness are given, it is not always clear how they have have been obtained. Moreover, we find that calculating a surface brightness depth based on the assumption that each pixel in a tidal feature must be detected with $S/N=3$, for example, leads to surface brightness depths that are too bright (i.e. suggesting low surface brightness sensitivity). This is because the integrating properties of the human eye-brain combination mean that features can be visually detected at much lower $S/N$ levels per pixel, as long as they are spread over several resolution elements. Indeed, by visually examining a range of archival HST images used in the host galaxy studies listed in Table \ref{tab:comparison}, as well as our own ground-based images of type 2 quasar objects, we find that the faintest diffuse tidal features that are confidently detected have $S/N\sim$1--1.3 per pixel\footnote{Note, however, that the presence of such faint features would not necessarily have led to a disturbed classification in the online interface, where the ability to manipulate the contrast levels and image sizes was limited. Therefore, this criterion may be conservative, in the sense that it overestimates the true surface brightness depth of the observations (i.e. is quoted too faint) -- features may have needed to be brighter in reality to lead to particular classifications in the interface.}.

As examples, we consider two of the most sensitive recent HST studies of quasar hosts -- those of \citet{vill17} and \citet{marian19} -- along with the older studies of \citet{dun03} and \citet{benn08}. The \citet{vill17} and \citet{marian19} studies have similar total exposure times of $\sim$1 orbit ($\sim$2400\,s) and both used the HST WFC3 instrument in the near-infrared with the F160W filter to study type 1 quasars. However, they study samples at different redshifts:  $z\sim0.6$ for \citet{vill17} and $z\sim2$ for \citet{marian19}. In contrast, the earlier study of \citet{dun03} used the HST WFPC2 instrument with a 1800\,s exposure time and the F675W filter to image quasars at $z\sim0.2$. Some of the objects in \citet{dun03} were later re-imaged by \citet{benn08} using much longer exposure times ($\sim$11,000s, 5 orbits) and the more sensitive HST ACS instrument with the F606W filter. Overall, we regard these four studies as being representative of the range of surface-brightness depths achieved in HST quasar host galaxy studies.

For each of the four studies we downloaded and (where necessary) combined the images available for each object in the HST archive. We then examined the images to identify the faintest tidal features that are confidently detected on visual inspection of the images, with full manipulation of the image contrast levels allowed (typically corresponding to $S/N\sim$1--1.3 in each pixel --- see above)\footnote{These features were not only
associated with the quasar hosts but also with interacting galaxies in the surrounding field.}. We then estimated the surface brightness of these faintest features in the rest-frame $V$-band, having corrected for cosmological surface brightness dimming and the necessary K-corrections. We regard the faintest-feature surface brightness as providing a realistic estimate of the effective surface brightness depth of each study. Details of the calculation of the surface brightness depths are given in Appendix \ref{sec:appendix}. Note that, because we do not know the spectral energy distributions (SEDs) and colours of the tidal features, for each study we repeated the calculation making two assumptions about the feature SED: in one case we assumed the SED of an elliptical galaxy, and in the other that of an Sbc galaxy. However, these different assumed SEDs made only small differences to the estimated surface brightness depths. 

The resulting effective surface brightness depths are compared in Figure \ref{fig:sb_comp} with the distribution of the rest-frame $V$-band surface brightness estimated for the brightest tidal feature detected for each of the disturbed type 2 quasar hosts. First, considering the comparison with \citet{dun03} and \citet{benn08}, it is notable that the \citet{dun03} observations would have missed $\sim$30--40 per cent of the brightest features detected in the type 2 quasar hosts, but the much deeper observations of \citet{benn08} would have detected them all. This demonstrates the effect of differences in the raw, observed-frame depth of the observations: relatively shallow ($\le$1 orbit) HST observations are likely to miss a substantial fraction of diffuse tidal features because, despite the reduced background light in the space-based observations, the high spatial resolution of HST means that the light is spread over many pixels on the detector. 

Second, considering the more recent 1-orbit HST studies of \citet{vill17} and \citet{marian19}, we see that the $z\sim0.6$ study of \citet{vill17} would have missed a similar proportion ($\sim$25--40 per cent)\footnote{This may represent a conservative lower limit on the proportion of features missed by the \citet{vill17} study, because it does not take into account the fact that \citet{vill17} re-binned their data to a finer pixel scale compared to the images in the archive (see Appendix~\ref{sec:appendix} for further details).} of the brightest tidal features to the \citet{dun03} study, whereas the $z\sim2$ \citet{marian19} study would have missed all of these features. This provides an illustration of the effect of $(1+z)^4$ cosmological surface brightness dimming, since these two studies used the same instrument, filter and  exposure time, but imaged samples of quasars at widely different redshifts.

Clearly, both observed-frame surface brightness depth and cosmological surface brightness dimming have an important effect on the rate of detection of tidal features in quasar host galaxies. However, other factors are  also likely to be significant. In particular, some studies focus on samples of type 1 quasars for which it is necessary to subtract the point spread function (PSF) of the bright nuclear point source in order to detect features close to nucleus. Typically this is important at radial distances $<$5 -- 10\,kpc from the nuclei, depending on the redshift and brightness of the quasar. Studies of high redshift quasar samples are particularly badly affected because their quasars tend to be brighter relative to the underlying host galaxies than those of lower redshift cases. The subtraction of the quasar PSF is never perfect, and may lead to residual cosmetic features in the subtracted image. Moreover, even with perfect subtraction, the $S/N$ will be reduced for features in the central regions of the host galaxies, due to the large photon counts and associated noise across the brighter parts of the quasar PSF. These effects make it more challenging to detect signs of morphological disturbance in the bulge regions of type 1 quasars than in their type 2 counterparts. On the other hand, it is unlikely that they could fully explain the differences between our results for type 2 quasar objects and some of the literature results on type 1 quasars, since most of the tidal features we detect are at sufficiently large radial distances from the nucleus ($>$10\,kpc) that they would not be significantly affected by the PSF subtraction and noise.

The classification methodologies of the various studies also show considerable variation. For example, some studies (as here) use blind, randomised comparisons with control samples \citep[e.g.][]{vill17,marian19,marian20}, whereas in others the comparison is not blind or randomised \citep[e.g.][]{ram11,bess12}, and some studies make no comparison with control samples at all \citep[e.g.][]{urrutia08,greene09,zhao19,zhao21}. A further difference is that in certain studies full manipulation of the image contrast levels, magnification etc. is allowed when doing the classification \citep[e.g.][]{ram11,bess12},  but in others (as here) the ability to manipulate the images to reveal faint or near-nuclear features against the diffuse light of the underlying host galaxy may be limited. Furthermore, some studies may only accept the brightest, highest-surface-brightness features (e.g. those due to a recent major merger) as evidence for galaxy interactions, while others may accept more subtle signatures.

In summary, many factors are likely to contribute to the wide range of results in the literature regarding the rate of morphological disturbance in quasar host galaxies. However, of these factors, observational surface brightness depth and cosmological surface brightness dimming are probably the most important.

\section{Discussion}
\label{sec:discussion}

The large and highly significant difference that we find between the morphological disturbance rates of the type 2 quasar hosts ($65^{+6}_{-7}$ per cent) and their controls ($22^{+5}_{-4}$ per cent) by itself provides strong evidence that galaxy interactions are the dominant trigger for type 2 quasars in the local universe. In this section we discuss whether these results can be generalised to all types of quasars, and whether other triggering mechanisms might contribute at some level.

If we concentrate on the comparison with ground-based studies of low and intermediate-redshift samples that achieve a similar surface-brightness depth, it is notable that the overall rate of morphological disturbance that we find for the type 2 quasar host galaxies ($65^{+6}_{-7}$ per cent) is entirely consistent with the 75$^{+7}_{-12}$ per cent derived by \citet{bess12} for a sample of 20 type 2 quasars with intermediate redshifts $0.3 < z < 0.41$.

In terms of comparison with samples of radio-loud AGN, the disturbance rate for the type 2 quasar sources is significantly below the 96$^{+1}_{-8}$ per cent deduced for the radio-loud AGN with quasar-like luminosities from \citet{ram11}. This difference is likely to be due to the contrasting classification methods used by \citeauthor{ram11} (\citeyear{ram11}; see \citealt{pierce22} for discussion), coupled with the better average seeing conditions for the Gemini South observations used in their study. Indeed, we estimate a $68^{+8}_{-11}$ per cent disturbance rate for the 22 3CR radio-loud quasars with $L_{\rm [OIII]} > 10^{35}$\,W and $z < 0.3$ in \citet{pierce22}, which were observed with the INT/WFC under moderate seeing conditions and classified using the same methods as the current work. This is remarkably similar to the disturbance rate we measure for the type 2 quasar hosts. Moreover, the disturbance enhancement factor relative to control galaxies that we measure for the type 2 quasars is also consistent with that measured for the 3CR quasars (see Figure~\ref{fig:merger_frac}). Together, these results suggest that galaxy interactions are the dominant triggering mechanism not only for the predominantly lower-radio-power type 2 quasars, but also for radio-loud AGN with quasar-like [OIII] luminosities.

Considering merger stage, we find that a majority of the disturbed objects in our sample are observed in a pre-coalescence phase ($61^{+8}_{-9}$ per cent), in contrast with the idea that quasars are triggered at the peaks of gas-rich mergers and become visible post-coalescence as the circum-nuclear dust is dispersed \citep{sanders88}. This proportion of pre-coalescence systems is higher than measured for the radio-loud objects of quasar-like luminosity in the \citet[][]{ram11}
and \citet{pierce22} samples (44 per cent\footnote{This  includes two objects we would classify as having tidally interacting companion galaxies in our classification scheme, but that were not considered as pre-coalescence systems in \citet{ram11}.} and 15 per cent respectively), and the intermediate-redshift type 2 quasar objects in the \citet{bess12} sample (47 per cent). However, despite these differences, a significant fraction of pre-coalescence systems are found in all of these samples.  This demonstrates that the gas flows induced by galaxy interactions are capable of triggering quasar activity well before the nuclei coalesce: in all the pre-coalescence systems in our sample, the projected nuclear separations are $\gtrsim$5\,kpc, corresponding to dynamical timescales $t_{\rm dyn} \gtrsim 10^8$\,yr. Clearly, the triggering of AGN in a pre-merger phase is not solely the preserve of lower-luminosity AGN \citep[e.g.][]{ell11}, but also occurs at high AGN luminosities.

An important question concerns whether the type 2 quasar results can be generalised to all local quasars, including the type 1 objects. If the distinct optical spectral appearances of type 1 and type 2 quasars were solely due to differences in orientation coupled with the obscuring effects of the circum-nuclear torus, but the host galaxy properties of the two groups were the same on average, then it would be possible to generalise the results. However, such generalisation would not be appropriate if, because of observational selection effects,  the amount and distribution of obscuring dust in the host galaxies of the two types were different on average. For example, the additional dust and gas brought into the nuclear regions by galaxy mergers might lead to quasars triggered in such events being preferentially classified as type 2 objects. In this context, it is notable that recent comparisons between the cold ISM properties and star formation rates of large samples of local type 1 and type 2 quasars have failed to find significant differences that would suggest that one group is more likely to be associated with galaxy mergers than the other \citep{shangguan19,bernhard22}. While this supports generalisation of our results, further deep imaging of local type 1 quasars  -- potentially challenging because of PSF subtraction issues --- is clearly required for confirmation.  

We further emphasise that, while our results present strong evidence that galaxy interactions are the {\it dominant} trigger for type 2 quasars in the local universe, it is unlikely that they are the {\it sole} trigger, since a significant fraction of the undisturbed objects in our sample show no signs of morphological disturbance on close inspection down to the surface brightness limit of our survey. For example, both J1100+0846 and J1509+0434 appear as undisturbed barred spiral galaxies (see Figure \ref{fig:undisturbed}), without evidence for large-scale tidal features outside their prominent disk components. Detailed observations of the molecular (CO) gas kinematics of these two objects show that the bulk of their cold gas is undergoing the regular gravitational motions expected of disk galaxies (including bar-induced motions), along with evidence for AGN-driven outflows \citep{ramos22}. Both systems also appear to be relatively isolated in terms of their large-scale environments. Therefore, secular processes \citep[e.g. bar-driven radial gas flows and disc instabilities;][]{hq10,hb14} are more likely to be triggering the activity, at least in these two cases.

Finally, it is important to add the caveat that our results apply to quasars in the local universe, but the dominant triggering mechanism may change with redshift, as the properties of the galaxy population as a whole (e.g. gas fractions) evolve. At high redshifts, for example, the gaseous disks of late-type galaxies are likely to be more massive and have higher surface densities than locally, potentially enhancing the effectiveness of secular mechanisms such as disk instabilities in triggering quasar activity \citep{bournaud11,gabor13}. 

\section{Conclusions}
\label{sec:conclusions}

Our deep imaging observations of nearby type 2 quasars provide strong evidence that galaxy interactions are the dominant triggering mechanism for quasars in the local universe, consistent with the results for other samples of nearby radio-loud and radio-quiet quasars that have been observed to a similar surface-brightness depth. Much of the apparent ambiguity of the results in this field is likely to be due to differences in the surface brightness depths of the observations combined with the effects of cosmological surface brightness dimming. Clearly, it is important that these factors are given full consideration in future studies of quasar triggering.

Beyond the dominance of galaxy interactions, there appears to be a wide range of circumstances under which luminous, quasar-like AGN are triggered. Although our results indicate that the gas flows associated with galaxy interactions can provide sufficient mass infall rates to the central SMBH to trigger quasar activity even well before the two nuclei have coalesced, some objects are triggered in a post-coalescence phase. Moreover, a minority of our sample are disk galaxies that appear undisturbed in deep imaging observations. Therefore, secular processes may sometimes be capable of triggering quasar activity, even if this is not the dominant mechanism at low redshifts.

\section*{Acknowledgements}

We thank the anonymous referee for their constructive comments that have helped improve the quality of the paper.
We also thank Mischa Schirmer for advice and assistance
concerning image processing with \textsc{THELI}. JP acknowledges support from the Science and Technology Facilities Council (STFC) via grant ST/V000624/1. CT and JH also acknowledge support from STFC. YG is supported by US National Science Foundation grant 2009441. CO is supported
by the National Sciences and Engineering Research Council of
Canada (NSERC). CRA and GS acknowledge financial support from the European Union's Horizon
2020 research and innovation programme under Marie Sk\l odowska-Curie
grant agreement No 860744 (BiD4BESt). CRA and PSB acknowledge support from the project 
``Feeding and feedback in active galaxies'', with reference
PID2019-106027GB-C42, funded by MICINN-AEI/10.13039/501100011033. 
CRA acknowledges financial support from the Spanish Ministry of Science and Innovation (MICINN) through the Spanish State Research Agency, under Severo Ochoa Centres of Excellence Programme 2020-2023 (CEX2019-000920-S).
Based on observations taken with the NASA/ESA Hubble
Space Telescope, obtained at the Space Telescope Science Institute
(STScI), which is operated by AURA, Inc. for NASA
under contract NAS5-26555. The Isaac Newton Telescope is operated
on the island of La Palma by the Isaac Newton Group of
Telescopes in the Spanish Observatorio del Roque de los Muchachos
of the Instituto de Astrofísica de Canarias. For the purpose of open access, the authors have applied a Creative Commons Attribution (CC BY) licence to any Author Accepted Manuscript version arising.

\section*{Data Availability}

The reduced data underlying this article will be shared on reasonable request to the corresponding author, and the raw data are publicly available via the ING archive.



\bibliographystyle{mnras}
\bibliography{ref_master_2.bib,ref_new.bib} 

\begin{thebibliography}{}
\makeatletter
\relax
\def\mn@urlcharsother{\let\do\@makeother \do\$\do\&\do\#\do\^\do\_\do\%\do\~}
\def\mn@doi{\begingroup\mn@urlcharsother \@ifnextchar [ {\mn@doi@}
  {\mn@doi@[]}}
\def\mn@doi@[#1]#2{\def\@tempa{#1}\ifx\@tempa\@empty \href
  {http://dx.doi.org/#2} {doi:#2}\else \href {http://dx.doi.org/#2} {#1}\fi
  \endgroup}
\def\mn@eprint#1#2{\mn@eprint@#1:#2::\@nil}
\def\mn@eprint@arXiv#1{\href {http://arxiv.org/abs/#1} {{\tt arXiv:#1}}}
\def\mn@eprint@dblp#1{\href {http://dblp.uni-trier.de/rec/bibtex/#1.xml}
  {dblp:#1}}
\def\mn@eprint@#1:#2:#3:#4\@nil{\def\@tempa {#1}\def\@tempb {#2}\def\@tempc
  {#3}\ifx \@tempc \@empty \let \@tempc \@tempb \let \@tempb \@tempa \fi \ifx
  \@tempb \@empty \def\@tempb {arXiv}\fi \@ifundefined
  {mn@eprint@\@tempb}{\@tempb:\@tempc}{\expandafter \expandafter \csname
  mn@eprint@\@tempb\endcsname \expandafter{\@tempc}}}

\bibitem[\protect\citeauthoryear{{Bahcall}, {Kirhakos}, {Saxe}  \&
  {Schneider}}{{Bahcall} et~al.}{1997}]{bahcall97}
{Bahcall} J.~N.,  {Kirhakos} S.,  {Saxe} D.~H.,   {Schneider} D.~P.,  1997,
  \mn@doi [\apj] {10.1086/303926}, \href
  {https://ui.adsabs.harvard.edu/abs/1997ApJ...479..642B} {479, 642}

\bibitem[\protect\citeauthoryear{{Bennert}, {Canalizo}, {Jungwiert},
  {Stockton}, {Schweizer}, {Peng}  \& {Lacy}}{{Bennert} et~al.}{2008}]{benn08}
{Bennert} N.,  {Canalizo} G.,  {Jungwiert} B.,  {Stockton} A.,  {Schweizer} F.,
   {Peng} C.~Y.,   {Lacy} M.,  2008, \mn@doi [\apj] {10.1086/529068}, \href
  {https://ui.adsabs.harvard.edu/abs/2008ApJ...677..846B} {677, 846}

\bibitem[\protect\citeauthoryear{{Bernhard}, {Tadhunter}, {Pierce}, {Dicken},
  {Mullaney}, {Morganti}, {Ramos Almeida}  \& {Daddi}}{{Bernhard}
  et~al.}{2022}]{bernhard22}
{Bernhard} E.,  {Tadhunter} C.~N.,  {Pierce} J.~C.~S.,  {Dicken} D.,
  {Mullaney} J.~R.,  {Morganti} R.,  {Ramos Almeida} C.,   {Daddi} E.,  2022,
  \mn@doi [\mnras] {10.1093/mnras/stac474}, \href
  {https://ui.adsabs.harvard.edu/abs/2022MNRAS.512...86B} {512, 86}

\bibitem[\protect\citeauthoryear{{Bessiere}, {Tadhunter}, {Ramos Almeida}  \&
  {Villar Mart{\'\i}n}}{{Bessiere} et~al.}{2012}]{bess12}
{Bessiere} P.~S.,  {Tadhunter} C.~N.,  {Ramos Almeida} C.,   {Villar
  Mart{\'\i}n} M.,  2012, \mn@doi [\mnras] {10.1111/j.1365-2966.2012.21701.x},
  \href {https://ui.adsabs.harvard.edu/abs/2012MNRAS.426..276B} {426, 276}

\bibitem[\protect\citeauthoryear{{Blecha}, {Snyder}, {Satyapal}  \&
  {Ellison}}{{Blecha} et~al.}{2018}]{blecha18}
{Blecha} L.,  {Snyder} G.~F.,  {Satyapal} S.,   {Ellison} S.~L.,  2018, \mn@doi
  [\mnras] {10.1093/mnras/sty1274}, \href
  {https://ui.adsabs.harvard.edu/abs/2018MNRAS.478.3056B} {478, 3056}

\bibitem[\protect\citeauthoryear{{Bournaud}, {Dekel}, {Teyssier}, {Cacciato},
  {Daddi}, {Juneau}  \& {Shankar}}{{Bournaud} et~al.}{2011}]{bournaud11}
{Bournaud} F.,  {Dekel} A.,  {Teyssier} R.,  {Cacciato} M.,  {Daddi} E.,
  {Juneau} S.,   {Shankar} F.,  2011, \mn@doi [\apjl]
  {10.1088/2041-8205/741/2/L33}, \href
  {https://ui.adsabs.harvard.edu/abs/2011ApJ...741L..33B} {741, L33}

\bibitem[\protect\citeauthoryear{{Bower}, {Benson}, {Malbon}, {Helly}, {Frenk},
  {Baugh}, {Cole}  \& {Lacey}}{{Bower} et~al.}{2006}]{bow06}
{Bower} R.~G.,  {Benson} A.~J.,  {Malbon} R.,  {Helly} J.~C.,  {Frenk} C.~S.,
  {Baugh} C.~M.,  {Cole} S.,   {Lacey} C.~G.,  2006, \mn@doi [\mnras]
  {10.1111/j.1365-2966.2006.10519.x}, \href
  {https://ui.adsabs.harvard.edu/abs/2006MNRAS.370..645B} {370, 645}

\bibitem[\protect\citeauthoryear{{Byrne-Mamahit}, {Hani}, {Ellison}, {Quai}  \&
  {Patton}}{{Byrne-Mamahit} et~al.}{2022}]{byrne_mamahit22}
{Byrne-Mamahit} S.,  {Hani} M.~H.,  {Ellison} S.~L.,  {Quai} S.,   {Patton}
  D.~R.,  2022, \mnras, submitted

\bibitem[\protect\citeauthoryear{{Cameron}}{{Cameron}}{2011}]{cameron11}
{Cameron} E.,  2011, \mn@doi [\pasa] {10.1071/AS10046}, \href
  {https://ui.adsabs.harvard.edu/abs/2011PASA...28..128C} {28, 128}

\bibitem[\protect\citeauthoryear{{Chambers} et~al.,}{{Chambers}
  et~al.}{2016}]{cham16}
{Chambers} K.~C.,  et~al., 2016, arXiv e-prints, \href
  {https://ui.adsabs.harvard.edu/abs/2016arXiv161205560C} {p. arXiv:1612.05560}

\bibitem[\protect\citeauthoryear{{Chiaberge}, {Gilli}, {Lotz}  \&
  {Norman}}{{Chiaberge} et~al.}{2015}]{chiaberge15}
{Chiaberge} M.,  {Gilli} R.,  {Lotz} J.~M.,   {Norman} C.,  2015, \mn@doi
  [\apj] {10.1088/0004-637X/806/2/147}, \href
  {https://ui.adsabs.harvard.edu/abs/2015ApJ...806..147C} {806, 147}

\bibitem[\protect\citeauthoryear{{Croton} et~al.,}{{Croton}
  et~al.}{2006}]{cro06}
{Croton} D.~J.,  et~al., 2006, \mn@doi [\mnras]
  {10.1111/j.1365-2966.2005.09675.x}, \href
  {https://ui.adsabs.harvard.edu/abs/2006MNRAS.365...11C} {365, 11}

\bibitem[\protect\citeauthoryear{{Di Matteo}, {Springel}  \& {Hernquist}}{{Di
  Matteo} et~al.}{2005}]{dm05}
{Di Matteo} T.,  {Springel} V.,   {Hernquist} L.,  2005, \mn@doi [\nat]
  {10.1038/nature03335}, \href
  {https://ui.adsabs.harvard.edu/abs/2005Natur.433..604D} {433, 604}

\bibitem[\protect\citeauthoryear{{Downes} \& {Solomon}}{{Downes} \&
  {Solomon}}{1998}]{downes98}
{Downes} D.,  {Solomon} P.~M.,  1998, \mn@doi [\apj] {10.1086/306339}, \href
  {https://ui.adsabs.harvard.edu/abs/1998ApJ...507..615D} {507, 615}

\bibitem[\protect\citeauthoryear{{Dunlop}, {McLure}, {Kukula}, {Baum}, {O'Dea}
  \& {Hughes}}{{Dunlop} et~al.}{2003}]{dun03}
{Dunlop} J.~S.,  {McLure} R.~J.,  {Kukula} M.~J.,  {Baum} S.~A.,  {O'Dea}
  C.~P.,   {Hughes} D.~H.,  2003, \mn@doi [\mnras]
  {10.1046/j.1365-8711.2003.06333.x}, \href
  {https://ui.adsabs.harvard.edu/abs/2003MNRAS.340.1095D} {340, 1095}

\bibitem[\protect\citeauthoryear{{Ellison}, {Patton}, {Mendel}  \&
  {Scudder}}{{Ellison} et~al.}{2011}]{ell11}
{Ellison} S.~L.,  {Patton} D.~R.,  {Mendel} J.~T.,   {Scudder} J.~M.,  2011,
  \mn@doi [\mnras] {10.1111/j.1365-2966.2011.19624.x}, \href
  {https://ui.adsabs.harvard.edu/abs/2011MNRAS.418.2043E} {418, 2043}

\bibitem[\protect\citeauthoryear{{Ellison}, {Brown}, {Catinella}  \&
  {Cortese}}{{Ellison} et~al.}{2019a}]{ellison19a}
{Ellison} S.~L.,  {Brown} T.,  {Catinella} B.,   {Cortese} L.,  2019a, \mn@doi
  [\mnras] {10.1093/mnras/sty3139}, \href
  {https://ui.adsabs.harvard.edu/abs/2019MNRAS.482.5694E} {482, 5694}

\bibitem[\protect\citeauthoryear{{Ellison}, {Viswanathan}, {Patton},
  {Bottrell}, {McConnachie}, {Gwyn}  \& {Cuillandre}}{{Ellison}
  et~al.}{2019b}]{ellison19b}
{Ellison} S.~L.,  {Viswanathan} A.,  {Patton} D.~R.,  {Bottrell} C.,
  {McConnachie} A.~W.,  {Gwyn} S.,   {Cuillandre} J.-C.,  2019b, \mn@doi
  [\mnras] {10.1093/mnras/stz1431}, \href
  {https://ui.adsabs.harvard.edu/abs/2019MNRAS.487.2491E} {487, 2491}

\bibitem[\protect\citeauthoryear{{Fabian}}{{Fabian}}{2012}]{fab12}
{Fabian} A.~C.,  2012, \mn@doi [\araa] {10.1146/annurev-astro-081811-125521},
  \href {https://ui.adsabs.harvard.edu/abs/2012ARA&A..50..455F} {50, 455}

\bibitem[\protect\citeauthoryear{{Fukugita}, {Ichikawa}, {Gunn}, {Doi},
  {Shimasaku}  \& {Schneider}}{{Fukugita} et~al.}{1996}]{fuk96}
{Fukugita} M.,  {Ichikawa} T.,  {Gunn} J.~E.,  {Doi} M.,  {Shimasaku} K.,
  {Schneider} D.~P.,  1996, \mn@doi [\aj] {10.1086/117915}, \href
  {https://ui.adsabs.harvard.edu/abs/1996AJ....111.1748F} {111, 1748}

\bibitem[\protect\citeauthoryear{{Gabor} \& {Bournaud}}{{Gabor} \&
  {Bournaud}}{2013}]{gabor13}
{Gabor} J.~M.,  {Bournaud} F.,  2013, \mn@doi [\mnras] {10.1093/mnras/stt1046},
  \href {https://ui.adsabs.harvard.edu/abs/2013MNRAS.434..606G} {434, 606}

\bibitem[\protect\citeauthoryear{{Greene}, {Zakamska}, {Liu}, {Barth}  \&
  {Ho}}{{Greene} et~al.}{2009}]{greene09}
{Greene} J.~E.,  {Zakamska} N.~L.,  {Liu} X.,  {Barth} A.~J.,   {Ho} L.~C.,
  2009, \mn@doi [\apj] {10.1088/0004-637X/702/1/441}, \href
  {https://ui.adsabs.harvard.edu/abs/2009ApJ...702..441G} {702, 441}

\bibitem[\protect\citeauthoryear{{Harrison}}{{Harrison}}{2017}]{harrison17}
{Harrison} C.~M.,  2017, \mn@doi [Nature Astronomy] {10.1038/s41550-017-0165},
  \href {https://ui.adsabs.harvard.edu/abs/2017NatAs...1E.165H} {1, 0165}

\bibitem[\protect\citeauthoryear{{Heckman} \& {Best}}{{Heckman} \&
  {Best}}{2014}]{hb14}
{Heckman} T.~M.,  {Best} P.~N.,  2014, \mn@doi [\araa]
  {10.1146/annurev-astro-081913-035722}, \href
  {https://ui.adsabs.harvard.edu/abs/2014ARA&A..52..589H} {52, 589}

\bibitem[\protect\citeauthoryear{{Heckman}, {Smith}, {Baum}, {van Breugel},
  {Miley}, {Illingworth}, {Bothun}  \& {Balick}}{{Heckman}
  et~al.}{1986}]{heck86}
{Heckman} T.~M.,  {Smith} E.~P.,  {Baum} S.~A.,  {van Breugel} W.~J.~M.,
  {Miley} G.~K.,  {Illingworth} G.~D.,  {Bothun} G.~D.,   {Balick} B.,  1986,
  \mn@doi [\apj] {10.1086/164793}, \href
  {https://ui.adsabs.harvard.edu/abs/1986ApJ...311..526H} {311, 526}

\bibitem[\protect\citeauthoryear{{Heckman}, {Kauffmann}, {Brinchmann},
  {Charlot}, {Tremonti}  \& {White}}{{Heckman} et~al.}{2004}]{heck04}
{Heckman} T.~M.,  {Kauffmann} G.,  {Brinchmann} J.,  {Charlot} S.,  {Tremonti}
  C.,   {White} S.~D.~M.,  2004, \mn@doi [\apj] {10.1086/422872}, \href
  {http://adsabs.harvard.edu/abs/2004ApJ...613..109H} {613, 109}

\bibitem[\protect\citeauthoryear{{Hopkins} \& {Quataert}}{{Hopkins} \&
  {Quataert}}{2010}]{hq10}
{Hopkins} P.~F.,  {Quataert} E.,  2010, \mn@doi [\mnras]
  {10.1111/j.1365-2966.2010.17064.x}, \href
  {https://ui.adsabs.harvard.edu/abs/2010MNRAS.407.1529H} {407, 1529}

\bibitem[\protect\citeauthoryear{{Hopkins}, {Hernquist}, {Cox}, {Di Matteo},
  {Robertson}  \& {Springel}}{{Hopkins} et~al.}{2005}]{hopkins05}
{Hopkins} P.~F.,  {Hernquist} L.,  {Cox} T.~J.,  {Di Matteo} T.,  {Robertson}
  B.,   {Springel} V.,  2005, \mn@doi [\apj] {10.1086/432463}, \href
  {https://ui.adsabs.harvard.edu/abs/2005ApJ...630..716H} {630, 716}

\bibitem[\protect\citeauthoryear{{Hopkins}, {Hernquist}, {Cox}  \&
  {Kere{\v{s}}}}{{Hopkins} et~al.}{2008}]{hopkins08}
{Hopkins} P.~F.,  {Hernquist} L.,  {Cox} T.~J.,   {Kere{\v{s}}} D.,  2008,
  \mn@doi [\apjs] {10.1086/524362}, \href
  {https://ui.adsabs.harvard.edu/abs/2008ApJS..175..356H} {175, 356}

\bibitem[\protect\citeauthoryear{{Johansson}, {Burkert}  \& {Naab}}{{Johansson}
  et~al.}{2009}]{johansson19}
{Johansson} P.~H.,  {Burkert} A.,   {Naab} T.,  2009, \mn@doi [\apjl]
  {10.1088/0004-637X/707/2/L184}, \href
  {https://ui.adsabs.harvard.edu/abs/2009ApJ...707L.184J} {707, L184}

\bibitem[\protect\citeauthoryear{{Kormendy} \& {Ho}}{{Kormendy} \&
  {Ho}}{2013}]{kormendy13}
{Kormendy} J.,  {Ho} L.~C.,  2013, \mn@doi [\araa]
  {10.1146/annurev-astro-082708-101811}, \href
  {https://ui.adsabs.harvard.edu/abs/2013ARA&A..51..511K} {51, 511}

\bibitem[\protect\citeauthoryear{{Koss} et~al.,}{{Koss} et~al.}{2021}]{koss21}
{Koss} M.~J.,  et~al., 2021, \mn@doi [\apjs] {10.3847/1538-4365/abcbfe}, \href
  {https://ui.adsabs.harvard.edu/abs/2021ApJS..252...29K} {252, 29}

\bibitem[\protect\citeauthoryear{{Lintott} et~al.,}{{Lintott}
  et~al.}{2008}]{lin08}
{Lintott} C.~J.,  et~al., 2008, \mn@doi [\mnras]
  {10.1111/j.1365-2966.2008.13689.x}, \href
  {http://adsabs.harvard.edu/abs/2008MNRAS.389.1179L} {389, 1179}

\bibitem[\protect\citeauthoryear{{Lintott} et~al.,}{{Lintott}
  et~al.}{2011}]{lin11}
{Lintott} C.,  et~al., 2011, \mn@doi [\mnras]
  {10.1111/j.1365-2966.2010.17432.x}, \href
  {http://adsabs.harvard.edu/abs/2011MNRAS.410..166L} {410, 166}

\bibitem[\protect\citeauthoryear{{Lotz}, {Jonsson}, {Cox}  \& {Primack}}{{Lotz}
  et~al.}{2008}]{lotz08}
{Lotz} J.~M.,  {Jonsson} P.,  {Cox} T.~J.,   {Primack} J.~R.,  2008, \mn@doi
  [\mnras] {10.1111/j.1365-2966.2008.14004.x}, \href
  {https://ui.adsabs.harvard.edu/abs/2008MNRAS.391.1137L} {391, 1137}

\bibitem[\protect\citeauthoryear{{Marian} et~al.,}{{Marian}
  et~al.}{2019}]{marian19}
{Marian} V.,  et~al., 2019, \mn@doi [\apj] {10.3847/1538-4357/ab385b}, \href
  {https://ui.adsabs.harvard.edu/abs/2019ApJ...882..141M} {882, 141}

\bibitem[\protect\citeauthoryear{{Marian} et~al.,}{{Marian}
  et~al.}{2020}]{marian20}
{Marian} V.,  et~al., 2020, \mn@doi [\apj] {10.3847/1538-4357/abbd3e}, \href
  {https://ui.adsabs.harvard.edu/abs/2020ApJ...904...79M} {904, 79}

\bibitem[\protect\citeauthoryear{{Martini}}{{Martini}}{2004}]{martini04}
{Martini} P.,  2004, in {Ho} L.~C.,  ed., Coevolution of Black Holes and
  Galaxies. p.~169 (\mn@eprint {arXiv} {astro-ph/0304009})

\bibitem[\protect\citeauthoryear{{Martini} \& {Weinberg}}{{Martini} \&
  {Weinberg}}{2001}]{martini01}
{Martini} P.,  {Weinberg} D.~H.,  2001, \mn@doi [\apj] {10.1086/318331}, \href
  {https://ui.adsabs.harvard.edu/abs/2001ApJ...547...12M} {547, 12}

\bibitem[\protect\citeauthoryear{{McLure}, {Kukula}, {Dunlop}, {Baum}, {O'Dea}
  \& {Hughes}}{{McLure} et~al.}{1999}]{mclure99}
{McLure} R.~J.,  {Kukula} M.~J.,  {Dunlop} J.~S.,  {Baum} S.~A.,  {O'Dea}
  C.~P.,   {Hughes} D.~H.,  1999, \mn@doi [\mnras]
  {10.1046/j.1365-8711.1999.02676.x}, \href
  {https://ui.adsabs.harvard.edu/abs/1999MNRAS.308..377M} {308, 377}

\bibitem[\protect\citeauthoryear{{Mechtley} et~al.,}{{Mechtley}
  et~al.}{2016}]{mechtley16}
{Mechtley} M.,  et~al., 2016, \mn@doi [\apj] {10.3847/0004-637X/830/2/156},
  \href {https://ui.adsabs.harvard.edu/abs/2016ApJ...830..156M} {830, 156}

\bibitem[\protect\citeauthoryear{{Pierce}, {Tadhunter}, {Ramos Almeida},
  {Bessiere}  \& {Rose}}{{Pierce} et~al.}{2019}]{pierce19}
{Pierce} J.~C.~S.,  {Tadhunter} C.~N.,  {Ramos Almeida} C.,  {Bessiere} P.~S.,
   {Rose} M.,  2019, \mn@doi [\mnras] {10.1093/mnras/stz1253}, \href
  {https://ui.adsabs.harvard.edu/abs/2019MNRAS.487.5490P} {487, 5490}

\bibitem[\protect\citeauthoryear{{Pierce} et~al.,}{{Pierce}
  et~al.}{2022}]{pierce22}
{Pierce} J.~C.~S.,  et~al., 2022, \mn@doi [\mnras] {10.1093/mnras/stab3231},
  \href {https://ui.adsabs.harvard.edu/abs/2021MNRAS.tmp.2947P} {510, 1163}

\bibitem[\protect\citeauthoryear{{Ramos Almeida}, {Tadhunter}, {Inskip},
  {Morganti}, {Holt}  \& {Dicken}}{{Ramos Almeida} et~al.}{2011}]{ram11}
{Ramos Almeida} C.,  {Tadhunter} C.~N.,  {Inskip} K.~J.,  {Morganti} R.,
  {Holt} J.,   {Dicken} D.,  2011, \mn@doi [\mnras]
  {10.1111/j.1365-2966.2010.17542.x}, \href
  {http://adsabs.harvard.edu/abs/2011MNRAS.410.1550R} {410, 1550}

\bibitem[\protect\citeauthoryear{{Ramos Almeida} et~al.,}{{Ramos Almeida}
  et~al.}{2012}]{ram12}
{Ramos Almeida} C.,  et~al., 2012, \mn@doi [\mnras]
  {10.1111/j.1365-2966.2011.19731.x}, \href
  {http://adsabs.harvard.edu/abs/2012MNRAS.419..687R} {419, 687}

\bibitem[\protect\citeauthoryear{{Ramos Almeida}, {Bessiere}, {Tadhunter},
  {Inskip}, {Morganti}, {Dicken}, {Gonz{\'a}lez-Serrano}  \& {Holt}}{{Ramos
  Almeida} et~al.}{2013}]{ram13}
{Ramos Almeida} C.,  {Bessiere} P.~S.,  {Tadhunter} C.~N.,  {Inskip} K.~J.,
  {Morganti} R.,  {Dicken} D.,  {Gonz{\'a}lez-Serrano} J.~I.,   {Holt} J.,
  2013, \mn@doi [\mnras] {10.1093/mnras/stt1595}, \href
  {http://adsabs.harvard.edu/abs/2013MNRAS.436..997R} {436, 997}

\bibitem[\protect\citeauthoryear{{Ramos Almeida} et~al.,}{{Ramos Almeida}
  et~al.}{2022}]{ramos22}
{Ramos Almeida} C.,  et~al., 2022, \mn@doi [\aap]
  {10.1051/0004-6361/202141906}, \href
  {https://ui.adsabs.harvard.edu/abs/2022A&A...658A.155R} {658, A155}

\bibitem[\protect\citeauthoryear{{Reyes} et~al.,}{{Reyes}
  et~al.}{2008}]{reyes08}
{Reyes} R.,  et~al., 2008, \mn@doi [\aj] {10.1088/0004-6256/136/6/2373}, \href
  {https://ui.adsabs.harvard.edu/abs/2008AJ....136.2373R} {136, 2373}

\bibitem[\protect\citeauthoryear{{Sanders} \& {Mirabel}}{{Sanders} \&
  {Mirabel}}{1996}]{sanders96}
{Sanders} D.~B.,  {Mirabel} I.~F.,  1996, \mn@doi [\araa]
  {10.1146/annurev.astro.34.1.749}, \href
  {https://ui.adsabs.harvard.edu/abs/1996ARA&A..34..749S} {34, 749}

\bibitem[\protect\citeauthoryear{{Sanders}, {Soifer}, {Elias}, {Madore},
  {Matthews}, {Neugebauer}  \& {Scoville}}{{Sanders} et~al.}{1988}]{sanders88}
{Sanders} D.~B.,  {Soifer} B.~T.,  {Elias} J.~H.,  {Madore} B.~F.,  {Matthews}
  K.,  {Neugebauer} G.,   {Scoville} N.~Z.,  1988, \mn@doi [\apj]
  {10.1086/165983}, \href
  {https://ui.adsabs.harvard.edu/abs/1988ApJ...325...74S} {325, 74}

\bibitem[\protect\citeauthoryear{{Schirmer}}{{Schirmer}}{2013}]{schirmer13}
{Schirmer} M.,  2013, \mn@doi [\apjs] {10.1088/0067-0049/209/2/21}, \href
  {https://ui.adsabs.harvard.edu/abs/2013ApJS..209...21S} {209, 21}

\bibitem[\protect\citeauthoryear{{Serber}, {Bahcall}, {M{\'e}nard}  \&
  {Richards}}{{Serber} et~al.}{2006}]{serber06}
{Serber} W.,  {Bahcall} N.,  {M{\'e}nard} B.,   {Richards} G.,  2006, \mn@doi
  [\apj] {10.1086/501443}, \href
  {https://ui.adsabs.harvard.edu/abs/2006ApJ...643...68S} {643, 68}

\bibitem[\protect\citeauthoryear{{Shangguan} \& {Ho}}{{Shangguan} \&
  {Ho}}{2019}]{shangguan19}
{Shangguan} J.,  {Ho} L.~C.,  2019, \mn@doi [\apj] {10.3847/1538-4357/ab0555},
  \href {https://ui.adsabs.harvard.edu/abs/2019ApJ...873...90S} {873, 90}

\bibitem[\protect\citeauthoryear{{Shangguan}, {Ho}  \& {Xie}}{{Shangguan}
  et~al.}{2018}]{shangguan18}
{Shangguan} J.,  {Ho} L.~C.,   {Xie} Y.,  2018, \mn@doi [\apj]
  {10.3847/1538-4357/aaa9be}, \href
  {https://ui.adsabs.harvard.edu/abs/2018ApJ...854..158S} {854, 158}

\bibitem[\protect\citeauthoryear{{Silk} \& {Rees}}{{Silk} \&
  {Rees}}{1998}]{silk98}
{Silk} J.,  {Rees} M.~J.,  1998, \aap, \href
  {https://ui.adsabs.harvard.edu/abs/1998A&A...331L...1S} {331, L1}

\bibitem[\protect\citeauthoryear{{Storchi-Bergmann} \&
  {Schnorr-M{\"u}ller}}{{Storchi-Bergmann} \&
  {Schnorr-M{\"u}ller}}{2019}]{storchi19}
{Storchi-Bergmann} T.,  {Schnorr-M{\"u}ller} A.,  2019, \mn@doi [Nature
  Astronomy] {10.1038/s41550-018-0611-0}, \href
  {https://ui.adsabs.harvard.edu/abs/2019NatAs...3...48S} {3, 48}

\bibitem[\protect\citeauthoryear{{Tadhunter}, {Dicken}, {Morganti}, {Konyves},
  {Ysard}, {Nesvadba}  \& {Almeida}}{{Tadhunter} et~al.}{2014}]{tad14b}
{Tadhunter} C.,  {Dicken} D.,  {Morganti} R.,  {Konyves} V.,  {Ysard} N.,
  {Nesvadba} N.,   {Almeida} C.~R.,  2014, \mn@doi [\mnras]
  {10.1093/mnrasl/slu135}, \href
  {https://ui.adsabs.harvard.edu/abs/2014MNRAS.445L..51T} {445, L51}

\bibitem[\protect\citeauthoryear{{Treister}, {Schawinski}, {Urry}  \&
  {Simmons}}{{Treister} et~al.}{2012}]{tre12}
{Treister} E.,  {Schawinski} K.,  {Urry} C.~M.,   {Simmons} B.~D.,  2012,
  \mn@doi [\apjl] {10.1088/2041-8205/758/2/L39}, \href
  {http://adsabs.harvard.edu/abs/2012ApJ...758L..39T} {758, L39}

\bibitem[\protect\citeauthoryear{{Urbano-Mayorgas} et~al.,}{{Urbano-Mayorgas}
  et~al.}{2019}]{urb19}
{Urbano-Mayorgas} J.~J.,  et~al., 2019, \mn@doi [\mnras]
  {10.1093/mnras/sty2910}, \href
  {http://adsabs.harvard.edu/abs/2019MNRAS.483.1829U} {483, 1829}

\bibitem[\protect\citeauthoryear{{Urrutia}, {Lacy}  \& {Becker}}{{Urrutia}
  et~al.}{2008}]{urrutia08}
{Urrutia} T.,  {Lacy} M.,   {Becker} R.~H.,  2008, \mn@doi [\apj]
  {10.1086/523959}, \href
  {https://ui.adsabs.harvard.edu/abs/2008ApJ...674...80U} {674, 80}

\bibitem[\protect\citeauthoryear{{Veilleux} et~al.,}{{Veilleux}
  et~al.}{2013}]{veilleux13}
{Veilleux} S.,  et~al., 2013, \mn@doi [\apj] {10.1088/0004-637X/776/1/27},
  \href {https://ui.adsabs.harvard.edu/abs/2013ApJ...776...27V} {776, 27}

\bibitem[\protect\citeauthoryear{{Villforth} et~al.,}{{Villforth}
  et~al.}{2017}]{vill17}
{Villforth} C.,  et~al., 2017, \mn@doi [\mnras] {10.1093/mnras/stw3037}, \href
  {https://ui.adsabs.harvard.edu/abs/2017MNRAS.466..812V} {466, 812}

\bibitem[\protect\citeauthoryear{{Westra}, {Geller}, {Kurtz}, {Fabricant}  \&
  {Dell'Antonio}}{{Westra} et~al.}{2010}]{westra10}
{Westra} E.,  {Geller} M.~J.,  {Kurtz} M.~J.,  {Fabricant} D.~G.,
  {Dell'Antonio} I.,  2010, \mn@doi [\pasp] {10.1086/657452}, \href
  {https://ui.adsabs.harvard.edu/abs/2010PASP..122.1258W} {122, 1258}

\bibitem[\protect\citeauthoryear{{Wilkinson}, {Ellison}, {Bottrell}, {Bickley},
  {Gwyn}, {Cuillandre}  \& {Wild}}{{Wilkinson} et~al.}{2022}]{wilkinson22}
{Wilkinson} S.,  {Ellison} S.~L.,  {Bottrell} C.,  {Bickley} R.~W.,  {Gwyn} S.,
   {Cuillandre} J.-C.,   {Wild} V.,  2022, arXiv e-prints, \href
  {https://ui.adsabs.harvard.edu/abs/2022arXiv220704152W} {p. arXiv:2207.04152}

\bibitem[\protect\citeauthoryear{{Wylezalek}, {Zakamska}, {Liu}  \&
  {Obied}}{{Wylezalek} et~al.}{2016}]{wyl16}
{Wylezalek} D.,  {Zakamska} N.~L.,  {Liu} G.,   {Obied} G.,  2016, \mn@doi
  [\mnras] {10.1093/mnras/stv3022}, \href
  {https://ui.adsabs.harvard.edu/abs/2016MNRAS.457..745W} {457, 745}

\bibitem[\protect\citeauthoryear{{Zakamska} et~al.,}{{Zakamska}
  et~al.}{2003}]{zak03}
{Zakamska} N.~L.,  et~al., 2003, \mn@doi [\aj] {10.1086/378610}, \href
  {https://ui.adsabs.harvard.edu/abs/2003AJ....126.2125Z} {126, 2125}

\bibitem[\protect\citeauthoryear{{Zhao}, {Ho}, {Zhao}, {Shangguan}  \&
  {Kim}}{{Zhao} et~al.}{2019}]{zhao19}
{Zhao} D.,  {Ho} L.~C.,  {Zhao} Y.,  {Shangguan} J.,   {Kim} M.,  2019, \mn@doi
  [\apj] {10.3847/1538-4357/ab1921}, \href
  {https://ui.adsabs.harvard.edu/abs/2019ApJ...877...52Z} {877, 52}

\bibitem[\protect\citeauthoryear{{Zhao}, {Ho}, {Shangguan}, {Kim}, {Zhao}  \&
  {Gao}}{{Zhao} et~al.}{2021}]{zhao21}
{Zhao} Y.,  {Ho} L.~C.,  {Shangguan} J.,  {Kim} M.,  {Zhao} D.,   {Gao} H.,
  2021, \mn@doi [\apj] {10.3847/1538-4357/abe8d4}, \href
  {https://ui.adsabs.harvard.edu/abs/2021ApJ...911...94Z} {911, 94}

\makeatother
\end{thebibliography}



\newpage
\appendix

\section{WHT images, seeing conditions and surface brightness estimates}
\label{sec:appendix}

\begin{figure*}
	\includegraphics[width=18cm]{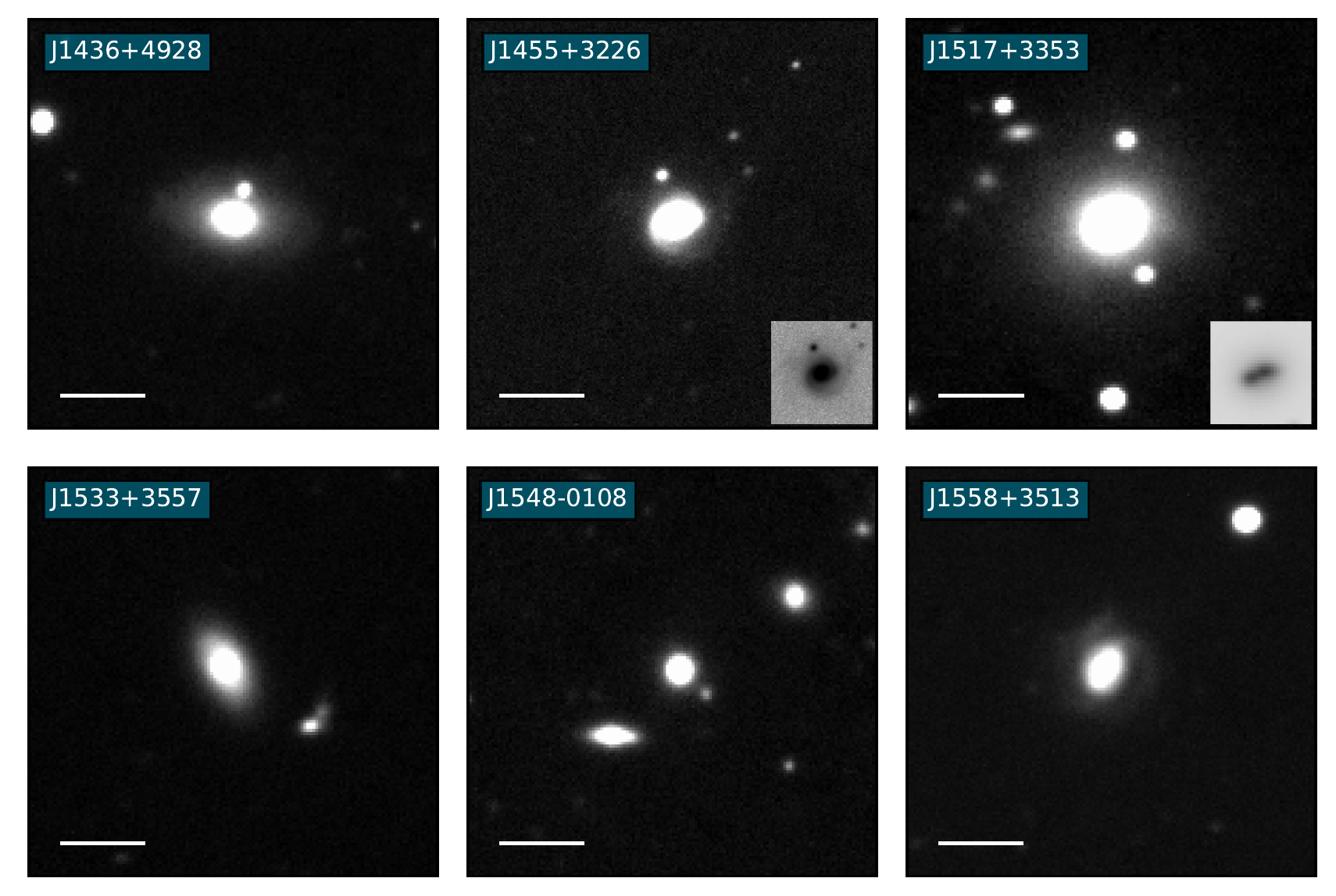}
    \caption{Thumbnail images of the 6 type 2 quasar objects observed with the WHT/PF-QHY. In each case north is to the top and east to the left, and the bar in the bottom left-hand corner indicates a distance of 20\,kpc at the redshift of the source. For J1455+3226  and J1517+3353 the insets show the central regions at a different contrast level to highlight near-nuclear structures. Note that the object $\sim$10\,kpc to the SW of J1548-0108 is a secondary nucleus/companion galaxy rather than a star, since its fitted FWHM is significantly larger than those of stars in the field.}
    \label{fig:wht}
\end{figure*}

The limiting surface brightnesses of the four studies  discussed in Section \ref{sec:comparison} and shown in Figure \ref{fig:sb_comp} were estimated by first examining images downloaded from the HST archive (co-added where necessary), and identifying the lowest surface brightness tidal features for each study that are confidently detected upon visual inspection of the images, allowing full manipulation of the image contrast levels.

In the case of the \citet{dun03} study, which used the F675W filter and the WFPC2 instrument, the surface brightnesses of the faintest features were measured directly from the images and converted into magnitudes using the zero points in the image headers. The F675W magnitudes were converted to F555W magnitudes using the transformation in the WFPC2 instrument handbook and assuming the $V-R_c$ colour for a $z=0.2$ elliptical galaxy from \citet{fuk96}. Then the magnitudes were converted to the observed-frame V-band using the $z=0.2$ elliptical galaxy $F_{555W}-V$ colour from \citet{fuk96}. Finally, the surface brightnesses were converted to the rest-frame V-band by correcting for $(1+z)^4$ surface-brightness dimming and applying the K-correction for $z=0.2$ from \citet{fuk96}. To assess the effect of assuming different colours for the features, the calculations were also repeated assuming Sbc galaxy colours.

For the \citet{benn08} study, which used the F606W filter and the ACS/WFC instrument, the steps were the same as those used for the estimates from the \cite{dun03} study, except that the F606W magnitudes were transformed directly to the observed-frame V band using $F_{F606W}-V$ colours for $z=0.2$ from \citet{fuk96}.

Both the \citet{vill17} and the \citet{marian19} studies used the near-IR F160W filter with the IR-channel of the HST WFC3 camera.  For these studies, lacking suitable colour transformations between the HST F160W and the observed-frame V-band for the redshifts of interest, the procedure used to calculate the limiting surface brightnesses was different from that used for the \citet{dun03} and \citet{benn08} studies: rather than directly measuring the surface brightnesses from the images, then transforming to the observed-frame V-band, the S/N of the faintest features was  estimated (typically $S/N\sim1.3$ found), then the HST WFC Exposure Time Calculator was used to estimate the observed-frame V-band surface brightness required to achieve the $S/N$ estimated for the faintest features. The last step involved correcting for $(1+z)^4$ surface-brightness dimming and K-correcting using the transformations for different 4000\AA\, break values\footnote{We assumed $D4000$ values of 1.9 and 1.3 for an elliptical galaxy and an Sbc galaxy respectively.} and redshifts from \cite{westra10}. Again, these calculations were done assuming both E and Sbc-galaxy colours.

Note that our limiting surface brightness estimates for the  \citet{vill17} and the \citet{marian19} studies are based on reduced images downloaded from the HST archive, which have the default WFC3 IR pixel scale of 0.13 arcsec pixel$^{-1}$. However, the results presented in both these studies are based on images re-sampled to a finer pixel scale of 0.06 arcsec pixel$^{-1}$. Given that this finer pixel scale will result in a lower $S/N$ (by a factor $\sim$2), the effective surface brightnesses of the images used in these studies could be brighter (i.e. less sensitive) by up to 0.75\,mag than the estimates shown in Figure \ref{fig:sb_comp}.

The WHT/PF-QHY images of the 6 type 2 quasar objects observed with this setup, as described in Section~\ref{sec:obs_and_class}, are shown in Figure~\ref{fig:wht}. Seeing FWHM estimates obtained from the INT/WFC and WHT/PF-QHY images of the type 2 quasar hosts are presented in Table~\ref{tab:obs_sb}, along with the V-band surface brightness measurements for all detected tidal features.

\begin{table*}
\centering
\caption{Seeing conditions and measured V-band surface brightness for tidal features detected in the
type 2 quasar objects. Where more than one number is given for the seeing FWHM in column 2, the first number corresponds to that measured from INT/WFC observations, while the second number corresponds to that measured from WHT/PF-QHY observations; otherwise all the seeing FWHM estimates are derived from INT/WFC observations. The rest-frame V-band surface brightness estimates shown in the final column have been K-corrected and corrected for cosmological surface brightness dimming.}
\label{tab:obs_sb}
\begin{tabular}{m{0.2\textwidth}>{\centering}m{0.2\textwidth}>{\centering \arraybackslash}m{0.3\textwidth}} 
\hline
Abbreviated name  & Seeing FWHM & Tidal feature surface brightnesses \\
& (arcsec) & ($\mu_V$; mag arcsec$^{-2}$) \\
\hline
J0052-0115 &1.37 &-- \\ 
J0232-0811 &1.31 &-- \\
J0731+3926 &1.39 &-- \\
J0759+5050 &1.47 &-- \\
J0802+4643 &1.72 &-- \\
J0802+2552 &1.40 &22.62, 23.47, 23.75, 24.01, 27.05, 27.37 \\
J0805+2818 &1.97 &24.04, 24.12 \\
J0818+3604 &1.24 &24.17, 24.54, 25.71, 25.88, 26.10 \\
J0841+0101 &2.10 &24.85, 25.20 \\
J0858+3121 &1.56 &24.99 \\
J0915+3009 &1.37 &22.63, 22.96, 23.91 \\
J0939+3553 &0.93 &24.34, 25.18, 25.30, 25.36, 25.56 \\
J0945+1737 &1.84 &23.26, 24.45, 24.60, 25.52, 25.60 \\
J1010+0612 &1.75 &22.93, 25.92, 26.00 \\
J1015+0054 &1.83 &25.45, 25.90, 26.37 \\
J1016+0028 &1.73 &23.78, 24.18, 25.40 \\
J1034+6001 &1.45 &22.97, 23.27, 23.66, 24.29, 26.06 \\
J1036+0136 &1.26 &23.68, 23.88 \\
J1100+0846 &1.18 &-- \\
J1137+6120 &1.46 &-- \\
J1152+1016 &1.11 &24.00, 24.88, 25.28, 25.56 \\
J1157+3707 &1.15 &-- \\
J1200+3147 &1.27 &21.54, 23.08, 24.42, 24.61 \\
J1218+4706 &1.00 &23.38, 24.05, 24.65, 25.95, 25.97 \\
J1223+0806 &1.11 &-- \\
J1238+0927 &2.32 &24.35 \\
J1241+6140 &1.63 &23.90, 24.63, 24.95, 26.68 \\
J1244+6529 &1.83 &23.37, 24.43 \\
J1300+5454 &1.64 &25.99, 26.32, 26.33, 26.55 \\
J1316+4452 &1.38 &23.78, 24.35, 24.82, 25.16, 25.79, 26.29 \\
J1347+1217 &1.92 &23.65, 24.05, 24.30, 25.86 \\
J1356-0231 &2.18 &24.61, 24.73 \\
J1356+1026 &2.04 &23.33, 23.35, 24.52, 24.85, 24.91 \\
J1405+4026 &1.66 &25.70, 25.77, 25.91 \\
J1430+1339 &2.01 &24.33, 24.69, 25.27 \\
J1436+4928 &1.58/0.94 &24.91, 26.44 \\
J1437+3011 &1.91 &23.70, 24.99, 25.14 \\
J1440+5330 &2.08 &24.00, 24.56 \\
J1455+3226 &1.94/1.00 &25.49, 26.09 \\
J1509+0434 &1.54 &-- \\
J1517+3353 &1.66/1.18 &-- \\
J1533+3557 &1.72/1.15 &--\\
J1548-0108 &2.28/1.13 &-- \\
J1558+3513 &1.99/1.08 &-- \\
J1624+3344 &1.56 &24.27, 24.33 \\
J1653+2349 &1.45 &-- \\
J1713+5729 &1.34 &-- \\
J2154+1131 &2.23 &-- \\
		\hline
	\end{tabular}
\end{table*}


\bsp	
\label{lastpage}
\end{document}